\documentclass[fleqn,usenatbib]{mnras} 

\usepackage[T1]{fontenc}
\usepackage{ae,aecompl}

\usepackage{graphicx}	
\usepackage{epstopdf}	
\usepackage{amsmath}	
\usepackage{amssymb}	
\usepackage{orcidlink}
\usepackage{siunitx}

\defcitealias{pettini2004iii}{PP04}
\defcitealias{marino2013o3n2}{M13}
\defcitealias{kewley2002using}{KD02}
\defcitealias{zaritsky1994h}{Z94}
\defcitealias{rupke2010gas}{R10}

\newcommand{\OHratio}{$\rm12+\log(O/H)$}
\newcommand{\FeHratio}{$\rm[Fe/H]$}
\newcommand{\kms}{{\rm km\,s^{-1}}}

\title[NGC 2207 and IC 2163: SITELLE and GCD+]{Integral field spectroscopy and numerical simulations of the NGC~2207/IC~2163 system}

\author[C. Poitras et al.]
{Camille Poitras$^{1}$\thanks{E-mail: camille.poitras.2@ulaval.ca}\orcidlink{0009-0001-5367-3976},
René Pierre Martin$^{2}$\orcidlink{0000-0002-6741-8298},
Laurent Drissen$^{1}$\orcidlink{0000-0003-1278-2591},
Hugo Martel$^{1}$\orcidlink{0000-0003-2917-2538},
\newauthor
and Carmelle Robert$^{1}$\orcidlink{0000-0003-2344-6593} \\
$^{1}$D\'epartement de physique, de g\'enie physique et d'optique, Universit\'e Laval, Qu\'ebec (QC), G1V 0A6, Canada\\
$^{2}$Department of Physics and Astronomy, University of Hawai'i at Hilo, Hilo, HI 96720, USA\\
}
\date{Accepted 2026 January 05. Received 2025 December 21; in original form 2025 September 15.}

\pubyear{2020}

\begin{document}
\label{firstpage}
\pagerange{\pageref{firstpage}--\pageref{lastpage}}
\maketitle

\begin{abstract}
    We present integral field spectroscopy of the interacting galaxy system NGC~2207/IC~2163 obtained with the imaging Fourier Transform Spectrometer SITELLE. Approximately 1000 \ion{H}{ii} region complexes are detected in both galaxies and analyzed using their strong optical emission lines. Their properties were studied via BPT diagrams and their luminosity function. We conducted a detailed study of the distribution of oxygen abundance across the system using a series of strong-line O/H indicators and calibrations. Both galaxies exhibit negative galactocentric abundance gradients with a slope $\sim$$-$0.015~dex~kpc$^{-1}$. There are marginal signs of discontinuities in the O/H gradients with some indicators while no significant azimuthal variations are seen. A shallower slope in the \ion{H}{ii} region luminosity function between the arm and inter-arm regions in IC~2163 is observed, supporting previous conclusion that the star formation process in this galaxy eyelids has been altered during the interaction. The kinematics of the ionised gas reveal disturbed velocity fields, AGN-like features in the nucleus of NGC~2207, and elevated velocity dispersion in turbulent or feedback-driven regions. To interpret these findings, we modeled the collision using the numerical algorithm GCD+. The simulation reproduces key features of the system and demonstrates how close passages drive enhanced star formation and localized chemical enrichment. Finally, two dwarf galaxies in the field are found to have very similar systemic velocities as their larger counterparts, and could well play a minor role in the global interaction based on their morphology and position.
\end{abstract}




\section{Introduction}

    Interactions and collisions between galaxies are common events in the Universe and can strongly alter galaxy evolution \citep{lambas2012galaxy, conselice2021galaxy}. Colliding galactic systems also provide opportunities to study diverse physical phenomena that contribute to shaping and transforming galaxies. In particular, processes defining the morphology of bulges and disks, creating tidal tails and streams (which could be sites of dwarf galaxies formation), and triggering starbursts and AGN activity can be investigated in such interactions \citep{anderson2013galactic, lanz2013global,  shah2022investigating,  asada2024bursty}. Other galaxy properties within interacting systems, for instance the level of star formation (SF) activity, can be driven by the presence of gas ﬂows \citep{barnes1996transformations}, although recent simulations suggest that the global star-forming history of galaxies is not always strongly influenced by interactions \citep{li2025effect}. 
    
    Another interesting aspect of galaxy interactions is how the initial chemical composition across the interacting objects could be altered during the colliding process. When present, gas inﬂows and outﬂows are suspected to modify the global distribution of chemical elements in colliding objects. Simulations have indeed suggested that dilution by low-metallicity gas inflows during interactions can result in flattening or erasing radial abundance gradients or even create positive gradients in some cases \citep{rupke2010galaxy, perez2011chemical}. From an observational point of view, metallicity gradients with a positive slope have been found in certain interacting galaxies, a fact very rarely seen in nearby isolated objects \citep{sanchez2014characteristic}. More importantly, there is also evidence that global O/H gradients might be shallower in a sample of interacting pairs studied by \cite{kewley2010metallicity} and \citet[hereafter R10]{rupke2010gas}. A recent integral field spectroscopy study of the interacting pair of galaxies in Arp 82 suggests also that gas flows generated during the interaction have strongly flattened the initial abundance gradient in one of the galaxies, a conclusion supported by numerical simulations duplicating the collision \citep{karera2022interacting}. Earlier studies on a few other systems have supported the idea that the gas chemical distribution is altered by an interaction, either across entire galaxies or in specific components like disc outskirts or tidal tails (e.g., \citealt{rich2012integral, bresolin2012gas, rosa2014interaction, olave2015ngc}). Similarly, lower central abundances have also been reported in interacting galaxies (\citetalias{rupke2010gas}); simulations have duplicated these results \citep{rupke2010galaxy, torrey2012metallicity}. Recently, \cite{pan2025sdss} have investigated radial O/H gradients in a sample of MaNGA galaxies within systems at different stages of interaction. As expected, most interacting galaxies exhibit flatter O/H gradients although the flattening is not always consistent across all cases and appears to vary depending on the interaction intensity. These observations,  coupled with numerical simulations, all suggest that gas ﬂows induced during interactions can indeed significantly alter the distribution of the chemical composition on a global scale. However, the resulting impact on the gas-phase metallicity might not be uniform and might strongly depend on the level and time evolution of the interaction. 
    
    The possibility of a galaxy interaction generating large-scale mixing and altering the distribution of the chemical composition in galaxies can be thoroughly studied in the well-known pair of spiral galaxies, NGC~2207 and IC~2163. This system ($D = 35$~Mpc, $1\arcsec = 170$~pc, $v = 2740$~km~s$^{-1}$) is undergoing a grazing collision and has been extensively studied from the X-rays to cm wavelengths. The larger galaxy, NGC~2207, exhibits strong spiral arms and several extended tidal structures. Some particularly intense zones of SF are observed in the spiral arms, in particular one ``mini-starburst'' object, dubbed {\it Feature~i}, in the western arm, well studied by \cite{elmegreen2006spitzer} and \cite{kaufman-ocular}. NGC~2207 has been the host of two recent core-collapse supernovae,  SN2010jp in the southern tidal tail \citep{corgan22-SN2010jp} and ASASSN-19kz  in the bright outer arm north of the nucleus \citep{strader-asassn-19kz}; this is consistent with the high star formation rate (SFR) measured in all wavebands from X-rays to radio \citep{mineo2014comprehensive,robin-sfr-uvit}. The smaller galaxy, IC~2163, exhibits an ``ocular'' structure with some intense SF in the oval arms (eyelids), and a long tidal tail in the eastern direction. Observations of more than 200 molecular gas complexes in both objects revealed clear differences between them depending on their location within the system \citep{elmegreen2017alma}. For example, the highest average mass is found in the eyelids of IC~2163. \cite{elmegreen2006spitzer} estimated the global SFRs in the system, based on 8 and 24 $\mu$m {\it Spitzer} observations, to be 10.8 and 13.7~M$_\odot$~yr$^{-1}$, respectively. On the other hand, from their X-ray and multi-wavelength analysis, \cite{mineo2014comprehensive} estimated the integrated SFR of the entire system to be roughly 24~M$_\odot$~yr$^{-1}$. In the same context, using a multi-wavelength approach, \citet{Mineo2013} inferred stellar masses of $M_* = 1.2\times10^{11}$~M$_\odot$ for NGC~2207 and $M_* = 5.2\times10^{10}$~M$_\odot$ for IC~2163. \ion{H}{i} {\citep{elmegreen1995interactionI} and CO observations \citep{kaufman-ocular} show important kinematics anomalies in both systems, in particular large $\sim$100~km~s$^{-1}$ streaming motions along the IC~2163 eyelids, probably resulting in the formation of the large complexes of molecular clouds seen in these components. In addition, at nuclear scales, X-ray observations indicate that NGC~2207 hosts a low-luminosity active galactic nucleus (AGN~; e.g., \citealt{elmegreen2006spitzer, kaufman2012ngc, mineo2014comprehensive}).
    
    As part of a general survey on metallicity in interacting pairs, long-slit spectra of 43 \ion{H}{ii} regions have been obtained in the disk of NGC~2207 and 19 in IC~2163 by \citetalias{rupke2010gas}. As seen with several strongly interacting pairs in their sample, the global O/H gradients in NGC~2207 and IC~2163, (with slopes $\Delta$~=~$-$0.0124 and $-$0.0165~dex~kpc$^{-1}$, respectively) appear significantly shallower than those of the non-interacting galaxies in their control sample. As mentioned earlier, this could be explained by large gas inﬂows/outﬂows generated during the encounter, as seen in \ion{H}{i} observations \citep{elmegreen1995interactionI}. However, the small number of \ion{H}{ii} regions studied does not allow the analysis of features such as breaks in the radial gradients or statistical measurements of the O/H fluctuations between the two objects. Also, the \ion{H}{ii} region luminosity function, an important parameter for studying initial mass function (IMF) variations and star-formation efficiency has yet to be established and requires a much larger dataset. 

    The NGC~2207/IC~2163 colliding system has also been studied with simulations, successful in reproducing a good number of morphological structural details. According to these models, the interaction is in its initial phase, with the closest interaction to have occurred $\sim$240 Myr ago with both galaxies expecting to merge in the future \citep{struck2005grazing}. Both original galaxies have been significantly altered during the collision. Simulations suggest that tidal forces are induced in the plane of IC~2163, creating large-scale shocks that resulted in the eyelids structure and the long tidal tails that are observed \citep{kaufman2012ngc}. Models also reveal that forces nearly orthogonal to the plane of the larger spiral NGC~2207 have warped the disk and triggered zones of active formation along its spiral arms, and some of the southern extensions seen in deep images of the system.

    In this paper, we revisit the ionised gas content of the NGC~2207/IC~2163 system with integral field spectroscopy using the imaging Fourier transform spectrometer (iFTS) SITELLE, supplemented by numerical simulations of the interaction with a smoothed particle hydrodynamics (SPH) code. We focus on the global oxygen abundance distribution across both galaxies and make use of simulations to investigate the global mixing taking place within the collision.

    This article is organized as follows. In Section~\ref{sec:section2} we describe our observations, data reduction, and calibration. The detection, spectrophotometry and a detailed analysis of the \ion{H}{ii} regions and the oxygen abundance distribution are presented in Section~\ref{section:HIIComplexes}. Global kinematics of the interaction as well as local velocity dispersion are analyzed in Section~\ref{section:Kinematics}. A comparison between numerical simulations and the observations is presented in Section~\ref{section:NumericalSimulations}. We report the radial velocity measurements of a small elliptical galaxy and the discovery of a dwarf star-forming galaxy in the field of view, both possibly associated with the system, in Section~\ref{section:quadruple}. Finally, we summarize our results in Section~\ref{section:Conclusions}.
    
\section{Observations, Calibration and extraction of the data}
\label{sec:section2} 
    The NGC~2207/IC~2163 system was observed with the iFTS SITELLE at the Canada-France-Hawaii Telescope (CFHT) in January 2019. The $11\arcmin \times 11\arcmin$ field of view (FOV) of this instrument fully covers the entire area of the colliding pair, with square spatial pixels (spaxels) of $0.32\arcsec$ \citep{drissen2019sitelle}. We used the SN1, SN2, and SN3 filters, which are centered on the bright lines of \ion{H}{ii} regions: [\ion{O}{ii}]$\lambda\lambda$3727-3729 (unresolved), H$\beta$, [\ion{O}{iii}]$\lambda$4959, [\ion{O}{iii}]$\lambda$5007, [\ion{N}{ii}]$\lambda$6548, H$\alpha$, [\ion{N}{ii}]$\lambda$6583, [\ion{S}{ii}]$\lambda$6716, and [\ion{S}{ii}]$\lambda$6731. Details and specifications of the observations are summarized in Table \ref{tab:DetailsObservations}.
    
        \begin{table}
            \centering
            \caption{Observations of the NGC~2207/IC~2163 system with SITELLE.}
            \label{tab:DetailsObservations}
            \begin{tabular}{lccccc}
                \hline
                Filter & Wavelength Range & $R$ & Exp./step & Num. Steps \\
                \hline
                SN1 & 363–386 nm & 950 & 57 s & 171 \\
                SN2 & 482–513 nm & 950 & 50 s & 220 \\
                SN3 & 647–685 nm & 2900 & 18 s & 506 \\                           
                \hline
            \end{tabular}
            \par\vspace{1ex}
            \parbox{0.9\linewidth}{\small Note : $R$ indicates the spectral resolution.}
        \end{table}

    Data reduction and analysis were performed using \texttt{ORBS} and \texttt{ORCS}, SITELLE's dedicated data reduction and analysis pipelines. Details of the automated procedure as well as the sources of uncertainties can be found in \citet{Martin2021}. 
    
    Photometric calibration has been secured from images and data cubes of spectrophotometric standard stars (GD71, GD 108 and LDS 749B). Wavelength calibration was performed using a high spectral resolution laser data cube observed at the zenith with an additional correction using the night sky OH emission lines (in the SN3 filter) in the science data cubes following the procedure described in \citet{2018MNRAS.473.4130M}. The uncertainty on the velocity calibration across the entire field is estimated to be less than 2 km s$^{-1}$ and within the main galaxies to less than 0.5 km s$^{-1}$. The barycentric correction was also applied to the observed spectra.

    From the calibrated data cubes, ORCS produces maps of the continuum, emission line amplitude and flux, velocity and velocity dispersion, as well as their respective uncertainty. ORCS can also be used to extract integrated spectra and values of the above-mentioned properties for any given defined region. Deep frames are also produced by adding all individual raw interferograms from both detectors. We note that the natural units of Fourier transform spectrometer's data are wavenumbers (cm$^{-1}$), which are used throughout this paper.

    In order to ensure a precise spatial correspondent between the data cubes in the different filters, we aligned them using the brightest stars common to each FOV, excluding stars located at the very edge of the FOV to minimize potential effects of optical distortion. Each stellar centroid was determined by fitting a two-dimensional Gaussian profile, and the resulting translation and rotation offsets between cubes were then computed. The measured and applied offsets were very small (less then one pixel).

    Figure~\ref{fig:NGC 2207-FOV} shows the composite image of the NGC~2207/IC~2163 pair produced by combining the SN2 and SN3 deep frames, enhanced with the H$\alpha$ map. The two galaxy nuclei are separated by a projected distance of $\sim 14$~kpc. Some of the main morphological structures generated by the collision have been identified; among them, the eyelids in IC~2163 defining its peculiar oval shape, the eastern tidal tail with its complex distribution of star forming regions, and the southern extension, the eastern spiral arm and the northern ``clump'' in the larger NGC~2207. A dwarf elliptical galaxy at the tip of IC~2163's tidal tail ($\sim 19$~kpc projected from the center of IC~2163), as well as a small spiral-like galaxy to the north-west ($\sim 45$~kpc projected from the center of NGC~2207), both likely part of the interacting system (see Section \ref{section:quadruple}), are also identified. Another image of this field, designed to enhance the faint structures outside the main galaxies, is presented in Figure~\ref{fig:DeepStarless}: it shows the full extent of the structural deformations induced by the interaction in the outskirts of both galaxies.
    
        \begin{figure*}
            \includegraphics[width=7.0truein]{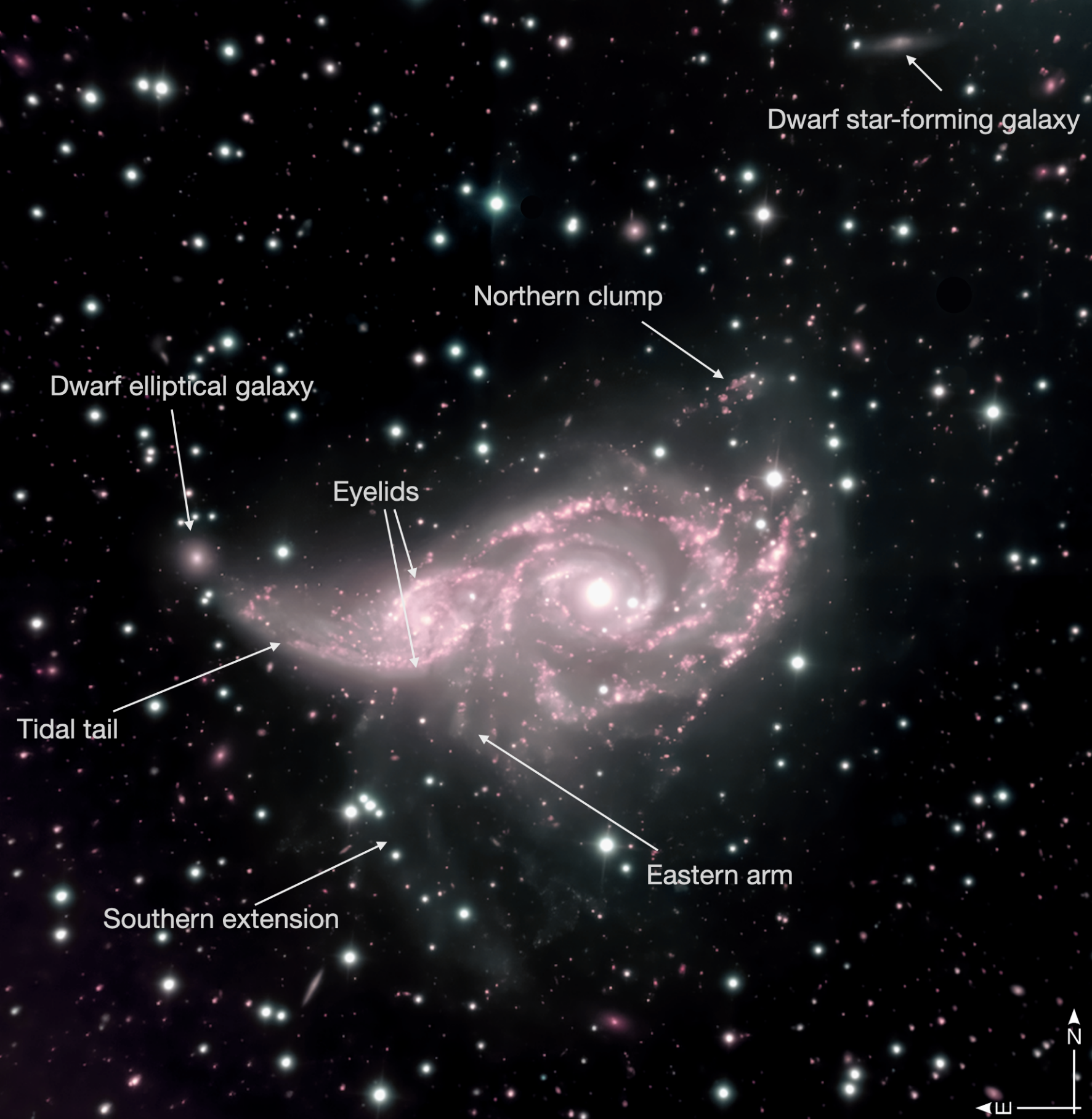}
            \caption{SITELLE image of the NGC~2207 (right) and IC~2163 (left) field, built from the deep SN2 and SN3 images combined with the H$\alpha$ map. The field of view is $8.7\arcmin \times 8.7\arcmin$. Arrows point to particular structures in the system and the two dwarf galaxies discussed in section \ref{section:quadruple}.}
            \label{fig:NGC 2207-FOV}
        \end{figure*}

\section{HII Region Complexes}
\label{section:HIIComplexes}
    In order to obtain integrated emission-line fluxes, and, consequently, the properties of \ion{H}{ii} region complexes in both galaxies, it is essential to first identify their positions and spatial boundaries. The key steps undertaken to achieve this identification are delineated in the following section. For more details regarding the identification and extraction of fluxes for the analysis of \ion{H}{ii} from SITELLE data, see \cite{rousseau2018ngc628}.

   \subsection{Detection and Measurements of Emission Regions}
   \label{section:Detection}
        Emission regions, i.e., mainly \ion{H}{ii} regions/complexes, are detected using a method well adapted to the spatial resolution, noise and multiple data cubes of SITELLE. Detailed insights into this procedure are provided in \cite{posternak_2025}. Herein, we present a concise overview of the primary steps involved in this methodology, with a specific emphasis on the parameters applied in the analysis of the NGC~2207 and IC~2163 system: 
        \begin{enumerate}
            \item \textit{Foreground Stars Removal and FOV Optimization} -
                   Prior to the detection process, we first minimized the impact of foreground stars in the proximity of NGC~2207 and IC~2163. This is necessary for preventing false positives and reducing background noise. We created masks to identify foreground stars across the FOV by comparing the positions of sources in the GAIA Data Release 3 Catalog \citep{brown2021gaia} and DAOStarfinder \citep{stetson1987daophot}, alongside a visual inspection of the SITELLE spectra. Concurrently, we reduced the FOV under consideration ($5.7\arcmin \times 5.4\arcmin$) to encompass only the galaxies of interest, thereby constraining the detection to emission regions within a logical proximity to our subjects of study.
            \vspace{4pt}
            \item \textit{Emission Peaks Detection} - 
                  Emission peaks are detected by applying a Gaussian smoothing to the Laplacian-transformed H$\alpha$ amplitude map with a standard deviation $\sigma_{\text{Lap}} = 1.5$ and a $3\times3$ spaxel detection box. These settings were tuned to reflect the prevailing observational seeing conditions, 
                  ensuring an effective separation of peaks in densely populated regions. To establish a quantifiable threshold T for peak detection, we calculated the value by incorporating considerations for the ambient spatial background level (BG) and spatial noise (Noi) characteristic of the H$\alpha$ amplitude, as per the framework proposed by \citet{posternak_2025}:
                        \begin{equation}
                            \text{T} = f_{\text{BG}}\times\text{BG} + \sigma_{\text{Noi}}\times \text{Noi} \;.
                        \end{equation}
                 In our assessment, each prospective emission peak was scrutinized against the background level measured over a larger 24×24 spaxel box. This analysis uses a background factor $f_{\text{BG}} = 0.95$ and a noise multiplier $\sigma_{\text{Noi}} = 1.8$. These values were selected by testing various combinations to ensure that peak detections correspond to the emission regions observable in the H$\alpha$ amplitude map. Peaks whose H$\alpha$ amplitude - averaged within the confines of the detection box - did not surpass the defined threshold were duly excluded from further analysis. The dimensions chosen for the background box are aimed at preserving the integrity of closely situated peaks of varying intensity levels while concurrently reducing the likelihood of erroneous detection due to the diffused ionised gas (DIG). The careful calibration to a $\sigma_{\text{Noi}}$ value ensures the recognition of emission peaks with an H$\alpha$ signal-to-noise (SNR) greater than 1. Using the above threshold parameters enabled the discernment of 1245 emission peaks, the spatial distribution of which is depicted in Figure~\ref{fig:DetectionCode}.
            \vspace{4pt}
            \item \textit{Zone of Influence Definition} - 
                  We defined a zone of influence (ZoI) around each identified peak by adopting spaxels that optimize the Amp(H$\alpha$) $r^{-2}$ ratio, where Amp(H$\alpha$) indicates the H$\alpha$ amplitude and $r$ denotes the radial distance to adjacent peaks. To accommodate all possible sizes for the ZoI, the parameter $r$ was constrained to a maximum value of $500$~pc (equivalent to ten spaxels). This distance is slightly greater than the maximum outer limit found for the emission regions of NGC~2207 and IC~2163 with SITELLE.
            \vspace{2pt}
            \item \textit{Final Emission Region Definition} - 
                 The emission region boundaries were established by fitting a 2D Gaussian plane profile to the ZoI's spaxels, using a straightforward Monte Carlo method to iteratively parameterize the Gaussian and the plane components representing background emissions. Specifically, we employed a Monte Carlo Markov Chain (MCMC) algorithm facilitated by 1000 walkers over 500 iterations for each emission region. Regions were defined by the intersection of the ZoI limit or the $3\sigma_{\text{Gau}}$ contour of the Gaussian fit, where $\sigma_{\text{Gau}}$ referred to the Gaussian's standard deviation along its major and minor axes, encapsulating the bulk of flux emitted by the region. The pixels location in the regions domains, as defined from the H$\alpha$ emission, was then reprojected in the other data cubes using their \textit{World Coordinate System} (WCS) information.
        \end{enumerate}
            \begin{figure*}
                \centering
                \includegraphics[width=\textwidth]{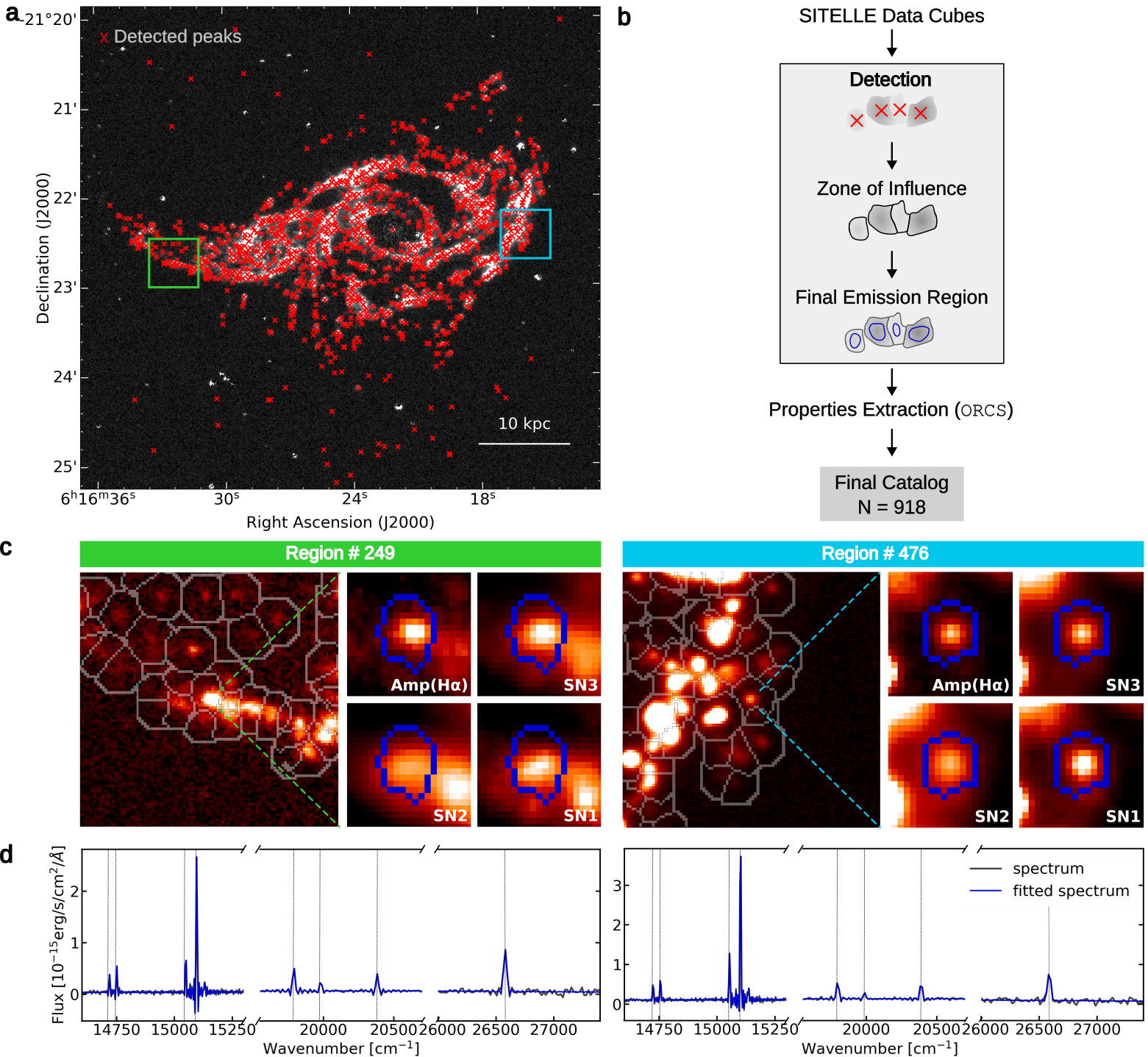}
                \caption{(a)~Emission regions initially detected (red crosses) on the H$\alpha$ amplitude map. (b)~Overview of the detection methodology for \ion{H}{ii} region complexes. It includes illustrations of the detection process, zone of influence and the final delineation of emission regions. (c)~Examples of detected emission regions in the tidal tail of IC~2163 (left) and a spiral arm of NGC~2207 (right) from the final catalog. Each example includes a broad view encompassing the zones of influence of neighboring regions on the H$\alpha$ amplitude map and a close-up view of an \ion{H}{ii} region complex displaying H$\alpha$ amplitude along with SN3, SN2, and SN1 deep frames. Different display levels are used across the map and deep frames to better highlight the emission distribution in each. (d)~Sky-subtracted spectra of these complexes are shown, highlighting the main fitted emission lines, from left to right, [\ion{S}{ii}]$\lambda$6731, [\ion{S}{ii}]$\lambda$6716, [\ion{N}{ii}]$\lambda$6583, H$\alpha$, [\ion{N}{ii}]$\lambda$6548, [\ion{O}{iii}]$\lambda$5007, [\ion{O}{iii}]$\lambda$4959, H$\beta$, and [\ion{O}{ii}]$\lambda$3727. Note that the oscillations near each line are not noise, but rather due to the natural instrument line shape, a sinc function.}
                \label{fig:DetectionCode}
            \end{figure*}
    
        Following the detection, the emission line fluxes, velocities, broadening and uncertainties of the final identified emission regions were extracted from their sky-subtracted integrated spectrum. This extraction method employed sinc profiles for the SN1 and SN2 data cubes and sincgauss profiles for the SN3 data cube using \texttt{ORCS}\footnote{SITELLE's instrument line shape is a sinc function; it is an excellent match to the low resolution ($R\simeq1000$) spectra provided by the SN1 and SN2 data cubes. A sincgauss, which is the convolution of a sinc with a gaussian, is preferred in the case of the higher resolution SN3 cube and allows to characterize the velocity dispersion along the line of sight \citep{Martin2016}.}. All lines in a data cube have been measured simultaneously. The sky subtraction was carried out by using a median sky spectrum derived from a region ($40 \times 40$ spaxels) located spatially away from the galaxy disk. Visual inspection of the spectra and the system revealed a lack of prominent Balmer absorption features, except in the region of the nucleus which also corresponds to one of the detected emission regions. The nucleus is very bright in all the main emission lines emitted within the SN3 filter band-pass, but was initially undetected in the H$\beta$ flux map. Since no other emission regions are located relatively close to the absorption observed near the AGN, we decided against subtracting any modeled stellar spectra from the extracted integrated spectra. 

        \subsubsection{Extinction Correction}
             We estimated an average dust extinction for each region using their integrated H$\alpha$/H$\beta$ ratio. Extinction corrections were then performed using \texttt{Pyneb} \citep{luridiana2015pyneb}, with the Cardelli extinction law \citep{cardelli1989relationship} and $R_V = 3.1$. A theoretical H$\alpha$/H$\beta$ flux ratio of 2.87 was assumed, consistent with Case B recombination \citep{osterbrock2006astrophysics} at a temperature of 10000~K. The extinction maps across the galaxies are presented in Figure~\ref{fig:Maps_av}. 
             
             We observe a clear difference in the extinction properties between the two galaxies. NGC~2207 exhibits a radial extinction gradient, with extinction values derived from the integrated fluxes of the final catalog ranging from $A_v \simeq 0.7$~mag near the center to $A_v \simeq 0.3$~mag toward the outskirts. These measurements are broadly consistent with those derived for NGC~2207 by \citet{berlind1997extinction} using the method of \citet{white1992direct} on visual and infrared images (spiral arms: $A_V \simeq 1$~mag; inter-arms: $A_V \simeq 0.5$~mag). Additionally, the northeastern spiral arm crosses over the central region of IC~2163, where we also measure relatively low extinction, around $A_v \simeq0.33$~mag. In contrast, IC~2163 - which lies partially behind NGC~2207 along the line-of-sight (LOS) - shows generally higher extinction in its central regions, with $A_v$ values ranging from approximately 0.9 to 1.7~mag. However, toward the end of its tidal tail, the extinction drops to approximately 0.3~mag. Overall, for the integrated fluxes from all emission-line regions in our final catalog (see Section~\ref{section:FinalCatalog}), we find an average extinction of $A_v = (0.51 \pm 0.02)$~mag. Furthermore, our result aligns well with the $A_v \simeq 0.55$~mag value reported by D.~S.~N. Rupke (priv. comm.) based on the data from \citetalias{rupke2010gas}. 
             
                \begin{figure}
                    \centering
                    \includegraphics[width=1.02\columnwidth]{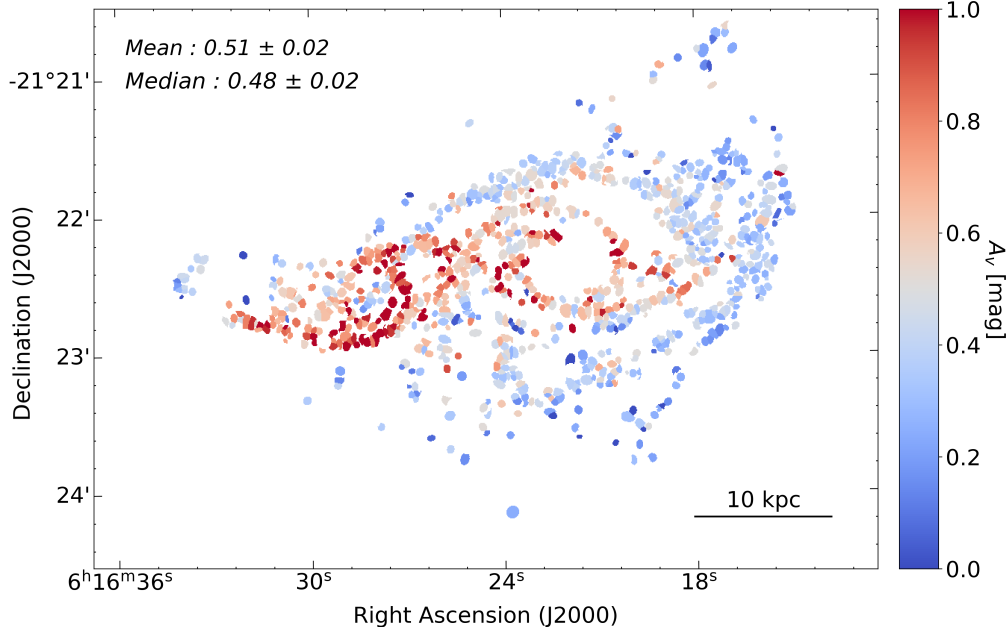}
                    \caption{The $A_v$ extinction maps for integrated fluxes from the final emission domains, derived from the H$\alpha$ to H$\beta$ ratio.}
                    \label{fig:Maps_av}
                \end{figure}

        \subsubsection{Comparison with Previous Measurements}
            For comparison purposes, we have analyzed the line ratios [\ion{N}{ii}]$\lambda$6583/H$\alpha$, [\ion{S}{ii}]$\lambda\lambda$6716-31/H$\alpha$ and [\ion{O}{iii}]$\lambda$5007/H$\beta$, and compared them to the corresponding values reported by \citetalias{rupke2010gas}, with thanks for providing the data. The \citetalias{rupke2010gas} regions corresponding to the SITELLE emission regions were identified based on the Euclidean distance in the sky plane. We find a strong correlation between the sets of ratios for regions within a distance of less than 3 spaxels (approximately our seeing limit), supported by a Pearson correlation coefficient $r \approx 1$, slopes consistently nearing one and intercepts approaching zero (see Figure~\ref{fig:ComparisonRupke}), accounting for uncertainties evaluated through a thousand random sampling simulations based on the uncertainties of the data. This agreement indicates that the flux calibration of our SITELLE data is sufficiently accurate for reliable line ratio measurement, which is central to the analyses presented in this study. Based on this agreement, we extend the comparison with \citetalias{rupke2010gas} to include oxygen abundances, as discussed later in Section~\ref{section:OxygenAbundance} for the final \ion{H}{ii} complexes defined in Section~\ref{section:Separation}. 

    \subsection{The Final HII Region Complexes Catalog}
    \label{section:FinalCatalog}
        Among the original sample resulting from the identification method described in the previous section, 327 regions (26~\%) displayed an integrated H$\alpha$ or H$\beta$ flux with a signal-to-noise ratio (SNR\footnote{For each line, we define the SNR as the ratio of total flux to its uncertainty, both determined by \texttt{ORCS}.}) inferior to 5. These regions were visually inspected: first using the H$\alpha$ map, followed by an examination of the three SN3 cube frames corresponding to their velocity range; in some cases, we also analyzed the integrated spectrum within the SN3 cube. Most of these regions are unambiguously discernible in H$\alpha$. We discarded 93 regions due to them being apparent false detections by the detection code or being too diffuse, likely part of the diffuse ionised medium, or DIG. Consequently, we retained 233 regions with a SNR~$< 5$ in the catalog, although they were excluded from further analysis.
        
        To classify ionization sources within galaxies and distinguish sources that are not related to SF processes, we rely on diagnostic line ratios in the Baldwin–Phillips–Terlevich (BPT) diagrams \citep{baldwin1981classification}. More specifically, we study two line-ratio diagnostics: the [\ion{O}{iii}]$\lambda$5007/H$\beta$ versus [\ion{N}{ii}]$\lambda$6583/H$\alpha$ and [\ion{O}{iii}]$\lambda$5007/H$\beta$ versus ([\ion{S}{ii}]$\lambda$6716 + [\ion{S}{ii}]$\lambda$6731)/H$\alpha$. Empirical boundaries established by \cite{kauffmann2003host} in the [\ion{O}{iii}]/H$\beta$ versus [\ion{N}{ii}]/H$\alpha$ diagram and theoretical limits set by \cite{kewley2001optical} in the [\ion{O}{iii}]/H$\beta$ versus [\ion{N}{ii}]/H$\alpha$ and [\ion{S}{ii}]/H$\alpha$ diagrams aid in distinguishing star-forming regions from other ionization sources within galaxies (e.g., AGNs, shocks). However, based on their finding of a strong correlation between gas phase velocity dispersion and optical emission line ratios in MaNGA data, \cite{law2021sdss} introduced new diagnostic lines that enhance the classification of ionised regions within galaxies. By adopting those, we identify only two regions that lie above these lines simultaneously in both [\ion{O}{iii}]/H$\beta$ versus [\ion{N}{ii}]/H$\alpha$ and [\ion{O}{iii}]/H$\beta$ versus [\ion{S}{ii}]/H$\alpha$ diagrams (see Figure~\ref{fig:BPTDiagrams}). In contrast, 35 other regions exceeded one of these lines in only one of the two diagrams. Figure~\ref{fig:BPTDiagrams} presents the BPT diagrams, colored based on the H$\alpha$ flux and point density, along with a fourth empirical boundary from \cite{schawinski2007observational} and [\ion{O}{iii}]/H$\beta$ versus [\ion{O}{iii}]/[\ion{O}{ii}]$\lambda$3727 diagram. The full sample is shown, as no significant differences are observed between the galaxies or between the arm and inter-arm regions (see Appendix~\ref{appendix:BPTDiagrams-Threshold}). We note that the increased dispersion of regions with lower H$\alpha$ flux is driven by larger uncertainties (see Figure~\ref{fig:BPTDiagrams-Threshold}). Consequently, all regions remain consistent with star-formation-dominated ionization and are retained for further analysis.

        Our final catalog includes a total of 1152 region complexes: 918 \ion{H}{ii} complexes considered for further analysis and 233 with a SNR~$< 5$ excluded from analysis. Notably, we are referring to complexes since, among them, 83~\% of the radii of the semi-major axes of the fitted Gaussian exceed 300 pc. Table \ref{tab:properties_complexes} provides the coordinates, integrated line fluxes, radial velocities, velocity dispersions, associated uncertainties, and final galaxy classification for each of the 918 analyzed complexes. The renaming 233 marginal detections are listed in Table \ref{tab:properties_marginal}, with the same properties reported when measurable, but without host galaxy assignments.

           \begin{figure*}
                \centering
                \includegraphics[width=\textwidth]{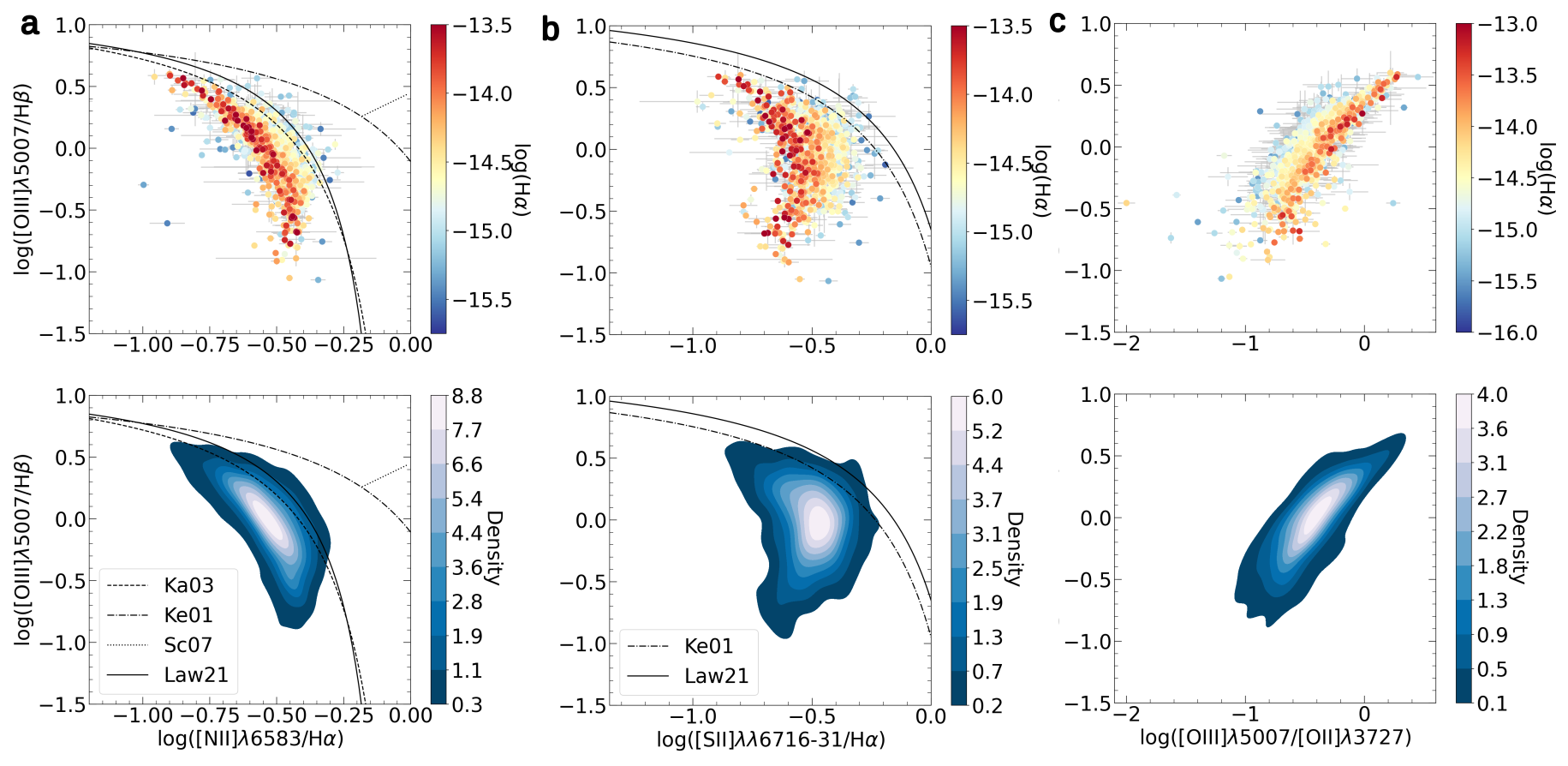}
                \caption{BPT diagrams for \ion{H}{ii} region complexes displaying the [\ion{O}{iii}]/H$\beta$ ratios as a function of (a)~[\ion{N}{ii}]/H$\alpha$, (b)~[\ion{S}{ii}]/H$\alpha$, and (c)~[\ion{O}{iii}]/[\ion{O}{ii}]. The color scheme in the top panel is determined by the logarithm of the H$\alpha$ flux, while in the bottom panel, the colors are based on point density. The boundaries are delineated by the equations of \protect\citet[dashed line]{kauffmann2003host}, \protect\citet[dash-dotted line]{kewley2001optical}, \protect\citet[dotted line]{schawinski2007observational}, and \protect\citet[solid line]{law2021sdss}.}
                \label{fig:BPTDiagrams}
            \end{figure*}

    \subsection{HII Region Complexes Physical Properties}
    \label{section:HIIComplexes2}
    
        \subsubsection{Separation of the \ion{H}{ii} Region Complexes for Both Galaxies}
        \label{section:Separation}
            The presence of a partial LOS overlap, with NGC~2207 positioned in the foreground, necessitates a spatial separation for a detailed analysis of each galaxy's \ion{H}{ii} region complexes and the evaluation of interaction effects. In their investigation of this interacting system, \cite{elmegreen1995interactionI} used the methodology proposed by \cite{brinks1984high} to disentangle \ion{H}{i} emission using data from the \textit{Very Large Array}, concluding that 80~\% of the gas in the overlap region is associated with NGC~2207, while 20~\% traces tidal debris from IC~2163. To study the distinctive characteristics of \ion{H}{ii} region complexes in each galaxy, including abundance gradients (Section~\ref{section:OxygenAbundance}), luminosity functions (Section~\ref{section:LuminosityFunctions}), and kinematics (Section~\ref{section:Kinematics}), we developed a novel approach to separate the complexes within their host galaxies.

            In the overlap zone, NGC~2207 intersects the western area of IC~2163's northern eyelid and the eastern part of its southern eyelid. Within this region, we have identified 240 \ion{H}{ii} regions complexes in our catalog that require determining their host galaxy. To achieve this, we have established five criteria that exhibit different values for each galaxy within the overlapping zone. These criteria are: the visual extinction $A_v$ and the emission line flux ratios log([\ion{N}{ii}]$\lambda$6583/H$\alpha$); log(([\ion{O}{iii}]$\lambda$5007/H$\beta$)/([\ion{N}{ii}]$\lambda$6583/H$\alpha$)); log([\ion{O}{iii}]$\lambda$5007/H$\beta$) and log([\ion{O}{ii}]$\lambda$3727/H$\beta$) ratios. These criteria were chosen for their natural gradient decrease from the galaxy's center, as well as their sensitivity to dust, metallicity and ionization state variations. We also considered velocities as a potential sixth criterion, since certain morphological structures display different velocities (see Section~\ref{section:Kinematics}). However, it was less indicative than the other five in the overlap zone, where multiple complexes shared the same range of values. 
        
            To achieve the separation of the complexes for both galaxies with these five criteria, our approach entailed two independent steps: (1) an initial manual examination leveraging criteria value distribution and observation of morphological structures; followed by (2) the implementation of an unsupervised machine learning algorithm, primarily based on $K$-means, which categorically clusters these regions according to their criteria values, devoid of morphological bias. 

            \textit{Manual Separation} - In the initial phase, we divided the \ion{H}{ii} region complexes visually, first through a conservative separation (leaving ambiguous cases unclassified) and then employing a non-conservative approach (assigning all complexes, even ambiguous ones) that assigns every \ion{H}{ii} region complex to one galaxy or the other. This procedure was conducted by visually inspecting each criterion, considering both the values and the morphological structures apparent (e.g., tidal tail, arm). From the non-conservative separation, complexes that were consistently categorized as belonging to the same galaxy across most criteria were assigned accordingly. Specifically, 74~\% of the complexes were always categorized under the same galaxy, 20~\% were categorized 4/5 times, and the remaining 6~\% had a 3/5 ratio, indicating a higher level of ambiguity and often excluded in the conservative separation. Ultimately, the non-conservative method yielded 128 complexes as part of IC~2163 and 112 complexes in NGC~2207.
                \begin{figure*}
                    \centering
                    \includegraphics[width=\textwidth]{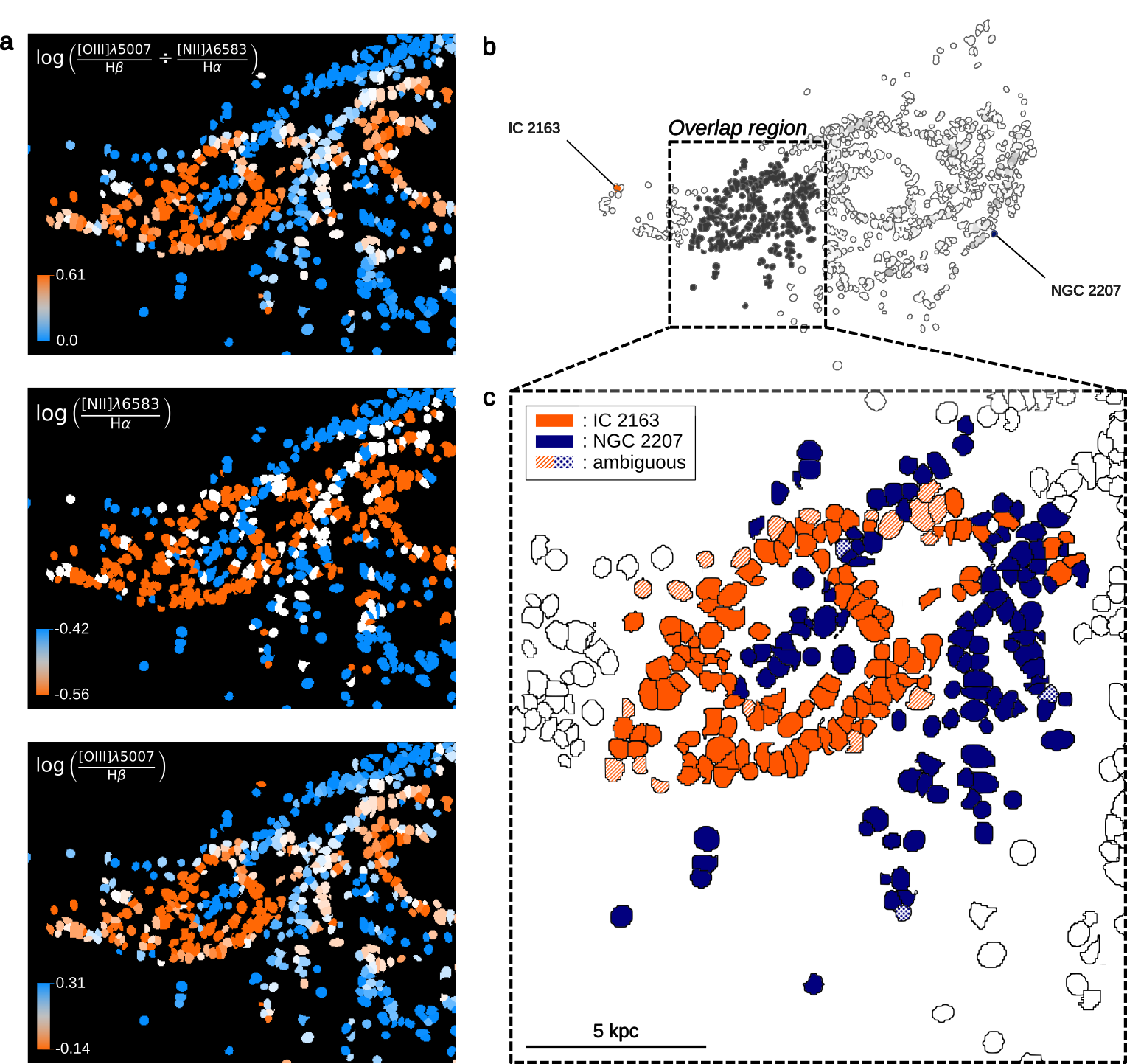}
                    \caption{Results of the separation of \ion{H}{ii} region complexes between NGC~2207 and IC~2163. (a)~Examples of three criteria used for the separation, with colors representing, top to bottom, the ratios log(([\ion{O}{iii}]$\lambda$5007/H$\beta$) / ([\ion{N}{ii}]$\lambda$6583/H$\alpha$)), log([\ion{N}{ii}]$\lambda$6583/H$\alpha$) and log([\ion{O}{iii}]$\lambda$5007/H$\beta$) from the integrated fluxes of the final emission domains. The colormap boundaries have been adjusted to distinguish different values in the overlap region. (b)~Contours delineating the domain of \ion{H}{ii} region complexes with the overlap region identified in darker shade. (c)~Zoomed-in view of the results in the overlap region depicted by the colored regions. The orange complexes are classified in IC~2163 by both methods, while the blue ones are classified in NGC~2207. The hatched regions represent complexes with ambiguous separations : the color corresponds to manual separation and the hatching indicates that the unsupervised machine learning classifies them in the other galaxy.}
                    \label{fig:SeparationComplexes}
                \end{figure*}

            \textit{Unsupervised Machine Learning Separation} - The second phase employed $K$-means clustering, an unsupervised method frequently utilized in astronomical studies (e.g., \citealt{turner2019reproducible, garcia2018machine}). $K$-means is an algorithm that partitions data into $K$ distinct clusters based on feature similarity, operating in a multidimensional space where each dimension corresponds to a criterion. Consequently, this phase focused on the criteria values while disregarding morphological structures. The $K$-means algorithm, with $K = 2$ representing the two galaxies, was executed over 1000 iterations to address the inherent stochastic nature of the method. A final decision on the arrangement of \ion{H}{ii} region complexes was then determined by hierarchical clustering of the $K$-means output, finalizing the separation process. As a result, 129 complexes were identified in IC~2163 and 111 complexes in NGC~2207.

            When comparing the outcomes of both methods, 91~\% of the \ion{H}{ii} region complexes in the overlap were classified in the same galaxy, leaving 22 complexes unclassified due to ambiguity. Figure~\ref{fig:SeparationComplexes} showcases the results of this separation along with three examples of criteria that facilitated this distinction. Complexes in ambiguity are primarily located in two areas: near the beginning of the tidal tail and partly in the outer periphery of the eyelids of IC~2163. These ambiguities were anticipated. Since IC~2163 mostly lies within the overlap region and the criteria exhibit a natural gradient, complexes farther from the center display different values that more resemble those of the complexes in the arm of NGC~2207. Furthermore, according to simulations (see Section~\ref{section:NumericalSimulations} and \citealt{struck2005grazing}), these complexes lie near the initial point of contact between the two galaxies. In this region, the physical conditions may have evolved due to the interaction, causing the properties of the complexes to change, thus complicating their attribution to a specific host. That said, the values of the separation criteria still tend to align slightly more closely with those observed in IC~2163.
            
            We performed a quick visual inspection of unpublished JWST+HST data, to verify if our separation appear consistent with what can be infer from images with finer details. The regions separated in our analysis appear clearly to belong to their respective galaxy in those images. However, the host galaxy remains challenging to identify for the ambiguous objects due to crowding effects or the difficulty in distinguishing entangled and crisscrossing morphological structures, even in the JWST+HST images.  
            
            In total, 729 complexes were identified within NGC~2207 and 167 complexes within IC~2163. This comprises 110 complexes from NGC~2207 and 108 complexes from IC~2163 located in the overlap zone. The 22 ambiguous complexes are included in the analysis of both galaxies in this study and will be identified as such.

        \subsubsection{Oxygen Abundances}
        \label{section:OxygenAbundance}

            As part of a general survey on metallicity in interacting pairs by \citetalias{rupke2010gas}, long-slit spectra of 43 \ion{H}{ii} regions in NGC~2207's disk and 19 in IC~2163's were analyzed. As observed for several strongly interacting pairs in their study, the global oxygen abundance gradients in both galaxies were found to be shallower than those of normal, isolated, spirals. However, the limited number of \ion{H}{ii} regions examined did not allow for an analysis of features such as breaks in the radial gradients or statistical assessments of the O/H fluctuations across both galaxies. With our number of detected \ion{H}{ii} region complexes being an order of magnitude larger compared to the \citetalias{rupke2010gas} study, we aim to further explore and provide a comprehensive analysis of the O/H  distribution within this interacting system. 
            
            To evaluate the gas-phase metallicity, we selected six widely-use calibrators from the literature. The use of this variety of calibrations will enhance the comparison of derived abundance values, thus avoiding potential systematic biases arising from issues such as inadequate extinction corrections or variations in ionization parameters of relative elemental abundances (\citealt{kewley2002using}, hereafter KD02; \citealt{morisset2025nebular}). The calibrations of interest in this study include N2 (\citealt{pettini2004iii}, hereafter PP04; \citealt{marino2013o3n2}, hereafter M13); O3N2 from \citetalias{pettini2004iii} and \citetalias{marino2013o3n2}; N2O2 \citepalias{kewley2002using}, based on [\ion{N}{ii}]$\lambda$6583/[\ion{O}{ii}]$\lambda\lambda$3727-3729 ratio; and R23 by \citet[hereafter Z94]{zaritsky1994h}, derived from ([\ion{O}{ii}]$\lambda\lambda$3727-29 + [\ion{O}{iii}]$\lambda$5007)/H$\beta$. For the N2 calibration by \citetalias{marino2013o3n2}, we applied their empirical calibration, valid for the $-1.6 < \log{(\text{N2})} < -0.2$, which encompasses all \ion{H}{ii} region complexes analyzed in this study. For the O3N2 diagnostic, we employed the theoretical calibration provided by \citetalias{pettini2004iii} for complexes falling within the $-1 < \log{\text{(O3N2)}} < 1.9$, and that of \citet{marino2013o3n2} for regions between $-1.1 < \log{(\text{O3N2})} < 1.7$. Most complexes fell within the specified intervals; however, 14 regions were excluded for \citetalias{pettini2004iii} and 21 regions were excluded for \citetalias{marino2013o3n2}. Concerning N2O2, we use the empirical calibration for $12 + \log{(\text{O/H})} > 8.6$ or $\log{(\text{N2O2})} > -0.97$, resulting in the exclusion of 8 complexes that do not fall within these ranges. Lastly, for R23, we implemented the theoretical calibration which includes regions within, generously, $8.4 < \log{(\text{O/H})} < 9.6$, retaining 873 complexes for this calibration.
    
            Figure~\ref{fig:AbundanceGradients} present the oxygen abundance for each galaxy as a function of the galactocentric distance $R$, with the aforementioned six calibrations. Galactocentric distances were computed using the same information on the three-dimensional alignment of each galaxy as per \citetalias{rupke2010gas} to ensure a valid comparison. These parameters were determined from numerical models \citep{elmegreen1995interactionII, elmegreen2000hubble}, with both galaxies having an inclination of 35$\degr$, along with position angles (PA) of 140$\degr$ and 128$\degr$ for NGC~2207 and IC~2163, respectively. Linear regressions were conducted using the least squares method, and the errors of the slope and intercept were estimated through a thousand random sampling simulations based on the uncertainties of the data.
         
                \begin{figure*}
                    \centering
                    \includegraphics[width=0.95\textwidth]{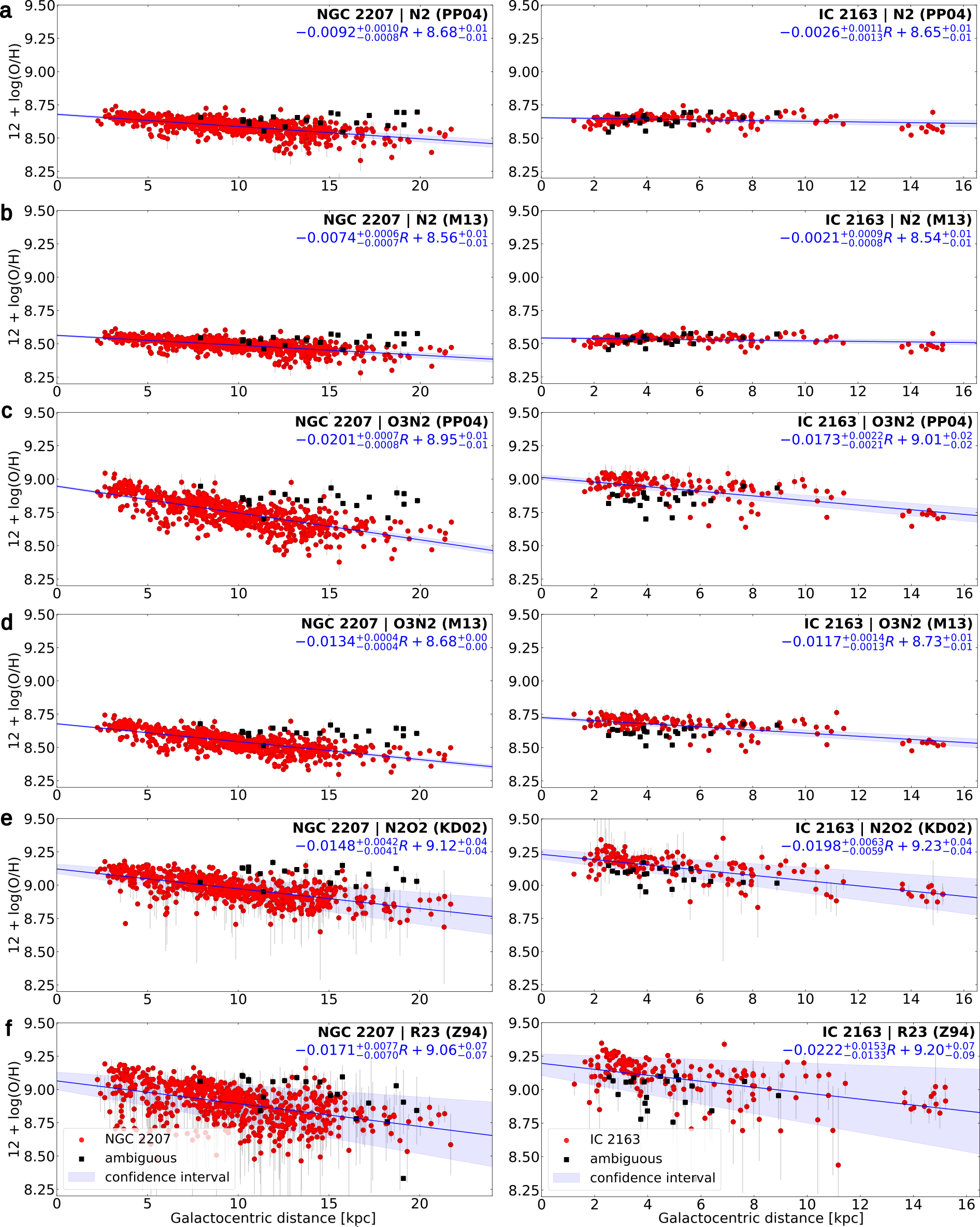}
                    \caption{Oxygen abundance profiles for \ion{H}{ii} region complexes for galactocentric distances within individual galaxies, specifically NGC~2207 (left pannel) and IC~2163 (right pannel). The various profiles are derived using the calibrations: (a)~N2 by \citetalias{pettini2004iii}, (b)~N2 by \citetalias{marino2013o3n2}, (c)~O3N2 by \citetalias{pettini2004iii}, (d)~O3N2 by \citetalias{marino2013o3n2}, (e)~N2O2 by \citetalias{kewley2002using}, and (f)~R23 by \citetalias{zaritsky1994h}. Linear regressions are performed using the red data points along with associated uncertainties to determine the confidence interval. Regression parameters are indicated at the top right. Black squares represent ambiguous complexes. As their host galaxy could not be determined, they have been included in the profile for each galaxy.}
                    \label{fig:AbundanceGradients}
                \end{figure*}

            For comparison, the metallicity gradient slopes derived here using the N2O2 indicator are consistent with those reported by \citetalias{rupke2010gas}. For NGC~2207, we measure a slope of $-0.0148^{+000042}_{-0.0041}$~dex~kpc$^{-1}$ compared to $-0.0124 \pm 0.0026$ in \citetalias{rupke2010gas}, while for IC~2163 we find $-0.0198^{+0.0063}_{-0.0059}$ compared to $-0.0165 \pm 0.0036$. In both cases, the measurements agree within the uncertainties.
            
            Several results on the O/H distribution in the NGC~2207/IC~2163 system can be observed from Figure~\ref{fig:AbundanceGradients}: 
            
            1) The central O/H value of both galaxies as extrapolated from the gradient linear regression is very similar for both galaxies for a given indicator; 
            
            2) The slopes of the global O/H gradient from the linear regression are similar between both galaxies for a given indicator, and the slopes are shallower than what is normally seen in typical spiral galaxies of similar masses. There is an important scatter in the slopes of the global O/H gradient among spiral galaxies, likely indicating different evolution histories, and the possible relationship with the galaxy mass (e.g., \citealt{zinchenko2021dependence}). For their survey, \citetalias{rupke2010gas} obtained an average slope of $\sim$ $-$0.05~dex~kpc$^{-1}$ for the global O/H gradient in their control sample (non-interacting galaxies), using the N2O2 indicator. This value is approximately 2-3 times steeper than the slopes measured for the NGC~2207/IC~2163 system in our study ($\sim -0.015$ to $-0.020$~dex~kpc$^{-1}$). Our results, based on a much larger sample of regions across both galaxies, are therefore consistent with the conclusions of \citetalias{rupke2010gas}.
            
            3) For some O/H indicators, the O/H radial gradients for both galaxies appear non-monotonic. For example, in the better sampled NGC~2207, a gradient flattening could be present at R > 15 kpc with O3N2. A positive slope within R < 8 kpc could be derived in IC~2163. According to \cite{sanchez2014characteristic}, flattening of the O/H gradient in the disk outskirts seems to be present in a large number of spiral galaxies, but they are not clear in interacting systems. However, these features are not present with all indicators and calibrations in our data, again raising the importance of using different methods and calibrators to study O/H distributions in galaxies due to large intrinsic uncertainties in these methods \citep{morisset2025nebular}. For example, the indicator O3N2 is more sensitive to the ionization parameter and the secondary nitrogen production, while N2O2 is more reliable but more affected by the correction for extinction \citep{kewley2002using}. Therefore, despite our large sample of regions and abundance values derived from different indicators and calibrations, there are no clear breaks in the radial abundance gradients for both galaxies.
            
            4) The ambiguous complexes, which have not been distinctly classified within a specific galaxy, exhibit generally higher oxygen abundances in NGC~2207 compared to the other complexes within that galaxy and the fitted gradients. In contrast, IC~2163 somewhat shows the opposite trend, with slightly lower oxygen abundances relative to its other complexes, except for N2, N2O2 and R23 as indicators. This distribution suggests that most of the ambiguous complexes are probably components of IC~2163. Due to our large data set, conclusions are not affected by keeping the ambiguous objects separate from the bulk of the other regions.  

            In addition to the radial profiles in abundances, it is worth exploring if possible large-scale azimuthal variations are seen in both systems. Such variations have been studied in several galaxies (see \citealt{bresolin2025signals}} and references therein) but remain elusive as large samples of \ion{H}{ii} are needed, while the intrinsic dispersion in abundances determined from the strong nebular line methods are significant. If present, azimuthal variations in the couple NGC~2207/IC~2163 could be due to rapid local enrichment from enhanced SF but also being generated at discontinuities in gas motions along important dynamical structures like the IC~2163 eyelids or the strong spiral arms in NGC~2207. However, quick enrichment could also be erased on a short time scale due to strong mixing induced by large streaming motions. In Figure~\ref{fig:AbundanceAz}, we present the sample of \ion{H}{ii} regions identified by their azimuthal location for both galaxies. The abundance gradients are presented using this color-coded identification for two different O/H indicators in both galaxies. No obvious azimuthal variations are seen, at least greater than the intrinsic $\sim$ 0.2-0.3 dex dispersion, for a given galactocentric distance range in both galaxies. The N2O2 indicator, in particular, less dependent on the ionization parameter and therefore with a smaller intrinsic dispersion, does not reveal large-scale variations. We will not push our analysis further here, but this is an interesting result for our simulations, as both galaxies seem to be currently relatively well mixed from a chemical composition point of view, at least within the intrinsic variations of the O/H indicators used.

            \begin{figure*}
                    \centering
                    \includegraphics[width=0.95\textwidth]{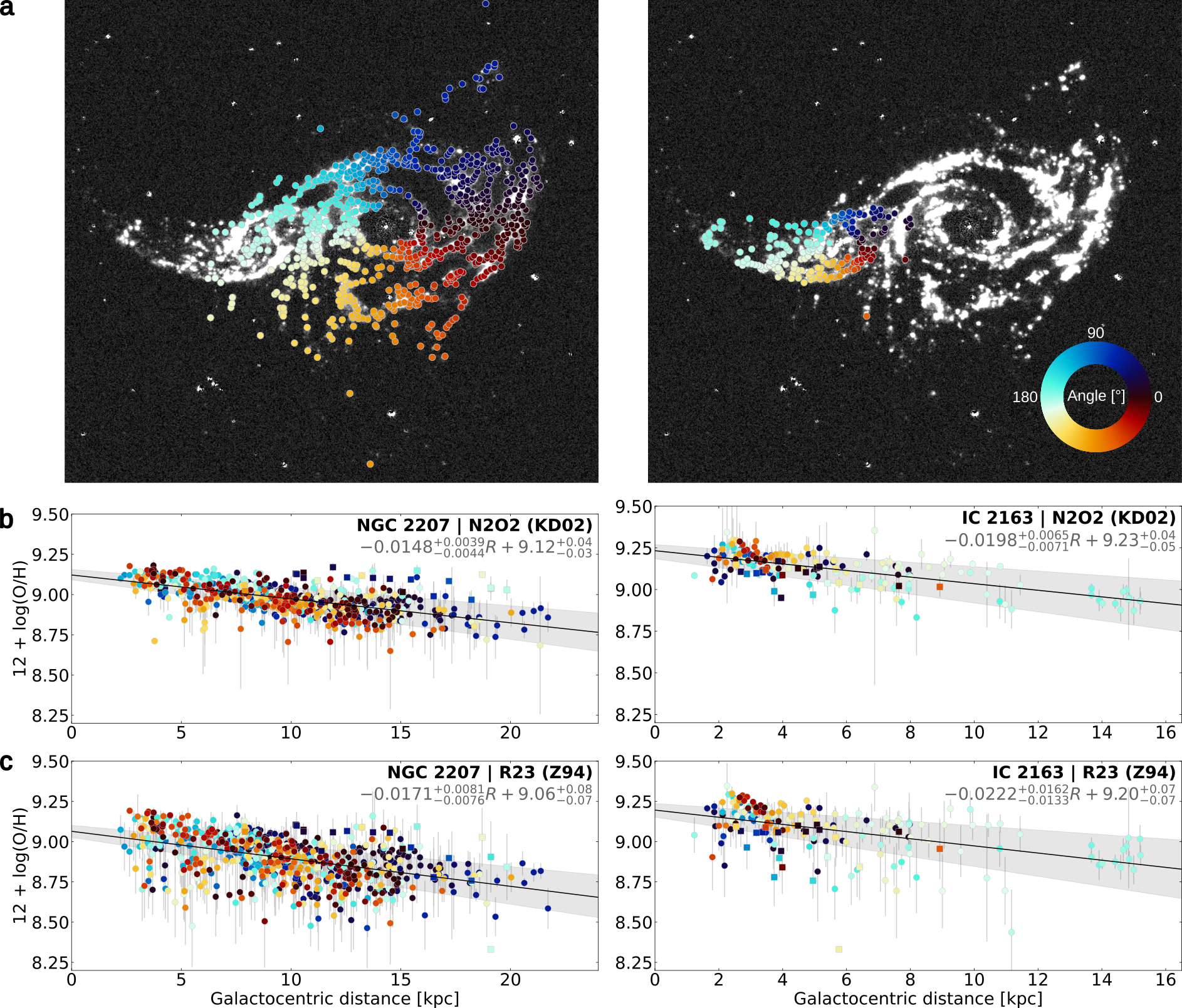}
                    \caption{Distribution of the \ion{H}{ii} region complexes in NGC~2207 (left panel) and IC~2163 (right panel), color-coded by azimuthal angle, as illustrated by the annular color bar (with $0\degr$ defined as the western direction). The lower panels show the abundance profiles from the N2O2 (\citetalias{kewley2002using}) and R23 (\citetalias{zaritsky1994h}) indicators for both galaxies. Each individual complexes is colored consistently with panel (a).}
                    \label{fig:AbundanceAz}
                \end{figure*}

        \subsubsection{Luminosity Functions}
        \label{section:LuminosityFunctions}
\begin{figure*}
                    \centering
                    \includegraphics[width=0.95\textwidth]{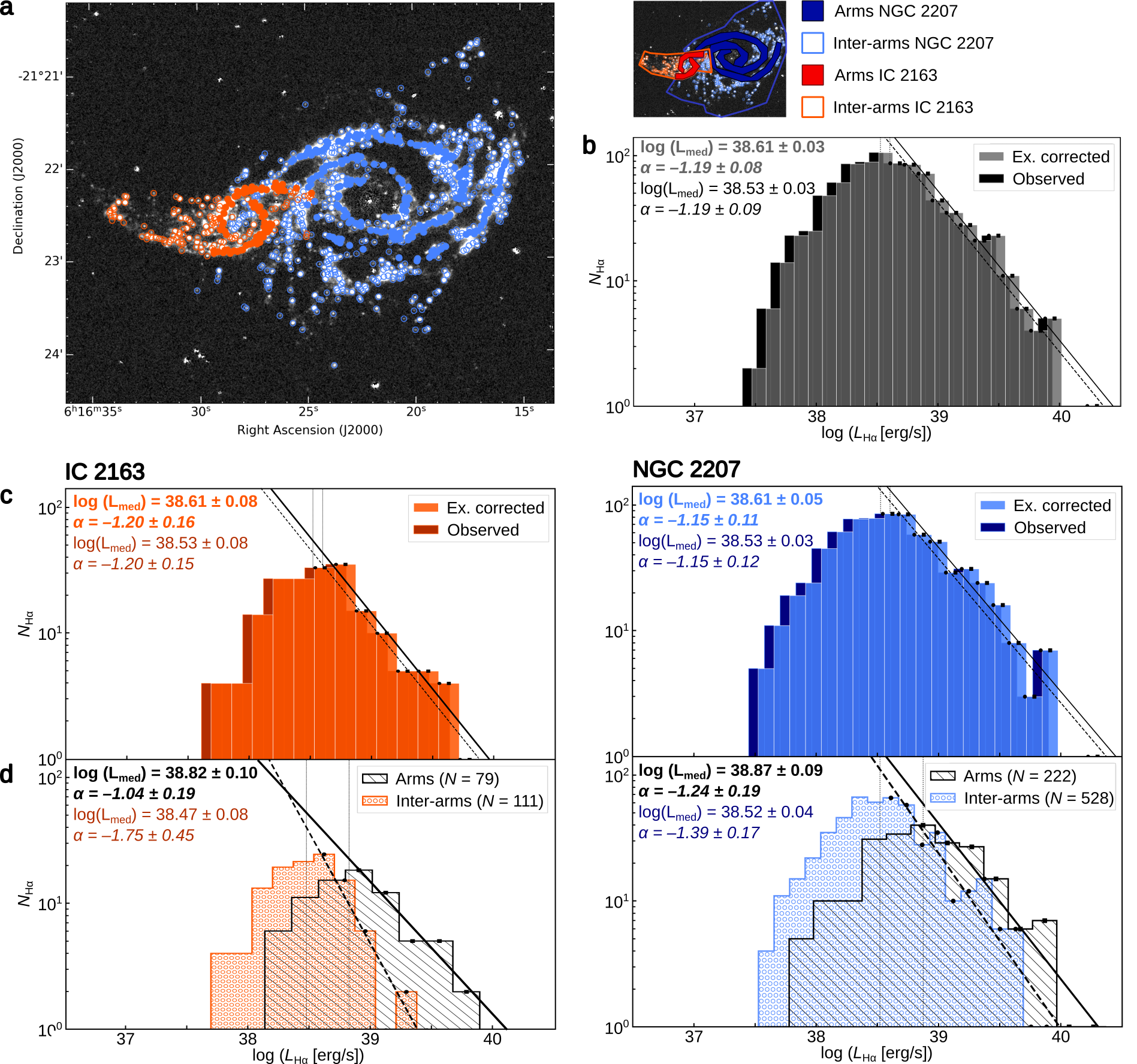}
                    \caption{(a)~\ion{H}{ii} region complexes divided and identified by galaxy and by morphological structures: spiral arm (filled circle) and inter-arm (empty circle). (b-c) Luminosity functions before and after extinction correction for (b)~the galaxy pair and (c)~IC~2163 and NGC~2207. The median $L_\text{med}$ and slopes $\alpha$ are indicated for both the corrected and uncorrected LFs, from top to bottom. (d)~Luminosity functions by morphological structure for each galaxy, with the estimated parameters $L_\text{min}$ and $\alpha$ identified. The legend includes, in parentheses, the number of complexes classified in each structure.}
                    \label{fig:LuminosityFunctions}
                \end{figure*}

            The H$\alpha$ flux and the extinction-corrected H$\alpha$ flux from the \ion{H}{ii} region complexes were converted into H$\alpha$ luminosity, using a distance of $D = 35$~Mpc for the galaxy pair (with $H_0 = 75$~km~s$^{-1}$~Mpc$^{-1}$, \citealt{elmegreen1995interactionI}). The resulting luminosity functions (LFs) for both the galaxy pair and individual galaxies, established post the separation of complexes (see Section~\ref{section:Separation}), are presented in Figure~\ref{fig:LuminosityFunctions}. It follows the relation
                \begin{equation}
                    N_{L_{\text{H}\alpha}}\text{d} L = A(L_{\text{H}\alpha})^\alpha\text{d} L \;,
                \end{equation}
            where $N_{L_{\text{H}\alpha}}\text{d}L$ is the number of complexes with a luminosity in the range $L$ to $L + \text{d}L$. The analysis also encompassed different morphological structures, including the spiral arms and inter-arms of each galaxy. To achieve this, the spiral arms were manually delineate. This identification was guided by the coherent, elongated morphology typically associated with spiral features. The separation between arms and inter-arms regions is indicated in the upper right panel of Figure~\ref{fig:LuminosityFunctions}. We tested whether adopting slightly broader arm definitions affect the resulting luminosity functions and found that, as long as the separation remained reasonable, the overall trends were robust. For the width of the bins, it was determined using the Freedman-Diaconis rule due to its suitability for our moderate dataset size, resulting in bins ranging from 0.12 to 0.24 in $\log(L_{\text{H}_\alpha} $[erg s$^{-1}$]).
            
            To fit the LFs, a completeness limit was determined by computing the median of each distribution. The uncertainty associated with $L_\text{med}$ was evaluated using a bootstrap method with a thousand iterations derived from the percentiles of bootstrapped medians. The slopes of the LFs $\alpha$ were then estimated using a linear least-squares regression in log-log space based on the centers of each bin. The uncertainty on the slope was also estimated using a bootstrap method with a thousand iterations where the standard deviation of the slopes obtained from resampling the data was calculated. 

            The distinction made in Figure~\ref{fig:LuminosityFunctions}d between the LFs of the arm versus the inter-arm regions reveals some interesting results. Within uncertainties, the general distribution suggests that the bulk of arm regions in both galaxies is more luminous by about log L(H$\alpha$) $\simeq$ 0.4 dex compared to the inter-arm regions. More importantly, the slope of the LF determined after the break at log L(H$\alpha$) $\simeq$ 38.5 (erg~s$^{-1}$) is close to identical for the arm and inter-arms regions in NGC~2207, within uncertainties, while noticing the smaller sample of arm regions. In IC 2163, the slope $\alpha$ = $-$1.75$\pm$0.45  for the inter-arm regions appears significantly steeper compared to the slope value $\alpha = -$1.04$\pm$0.19  for the arm regions. The relative uncertainties here are larger because the number of regions is smaller. If the difference in slope is real for both populations of regions in IC 2163, it represents a rare case where such a discrepancy has been observed in spiral galaxies \citep{rozas1995statistics}. For instance, a difference between the slopes of the arm vs inter-arm \ion{H}{ii} regions has been observed in the interacting galaxy M51 \citep{rand1992h, lee2011h}, and possibly for NGC 4321 \citep{cepa1990distribution}. If the steeper slope in the arm regions LF is real, it is tempting to suggest that the mass spectrum of molecular clouds forming the \ion{H}{ii} complexes in IC~2163 is shallower in the arms of the galaxy, in particular in the eyelids. That conclusion is supported by molecular gas observations obtained by \cite{elmegreen2017alma}: no difference is seen between the mass distribution of arm vs. inter-arm molecular complexes in NGC~2207 but the slope of the molecular cloud mass distribution in IC~2163 is steeper in the inter-arms. This behavior could be associated to the large star-formation activity in the eyelids generated by large-scale shocks during the encounter \citep{kaufman2012ngc}.

        \subsubsection{Integrated Star Formation Rate}
        \label{section:SFRs}

            The total H$\alpha$ flux for the pair\footnote{Measured within a circular aperture of 2.7 arcmin radius centered at R.A. 06h16m24.21s, Dec. -21°22\arcmin23.3\arcsec}, without any correction for extinction, is F(H$\alpha$)= 5.2 $\times$ 10$^{-12}$~erg~cm$^{-2}$~s$^{-1}$. Using the average extinction obtained from the integrated Balmer line decrement (Section~\ref{section:Separation}) leads to a corrected value of F(H$\alpha$)= 8.2 $\times$ 10$^{-12}$ erg~cm$^{-2}$~s$^{-1}$, corresponding to a luminosity of L(H$\alpha$)= 1.3 $\times$ 10$^{42}$~erg~s$^{-1}$ at a distance of 35~Mpc. Finally, using the conversion factor provided by \cite{kennicutt2012sfr} leads to a global SFR = 6.5~M$_\odot$~yr$^{-1}$. This value, derived from H$\alpha$, is lower by a factor of 2 than that obtained from the dust emission in the same area:  10.8 M$_\odot$~yr$^{-1}$ using the Spitzer 8$\mu$m flux and 13.7 M$_\odot$~yr$^{-1}$ from the 24$\mu$m flux \citep{elmegreen2006spitzer}. The SFR derived from X-ray measurements by \cite{mineo2014comprehensive}, also covering an area almost identical to ours, normalized to the distance used in this study, is even higher, at 18 M$_\odot$~yr$^{-1}$. Although these discrepancies are not abnormally high, several reasons might explain them:
            
        \begin{enumerate}
            \item The extinction has been underestimated. We have indeed applied an oversimplified average extinction correction (A$_V = 0.5$) to the integrated H$\alpha$ flux of the system. Studying a sample of 81 nearby galaxies, \cite{Kewley2002hair} found a correlation between the SFR determined from the raw (uncorrected for absorption) integrated H$\alpha$ flux and that determined from the far infrared (IRAS $60 - 100 \mu$m) flux; the slope of the relation (their Figure 1 and equation 8) is very steep, and a large scatter is observed. Scaling our measured uncorrected H$\alpha$ flux using this relation would lead to a SFR of 17 M$_\odot$~yr$^{-1}$.

            \item We have not applied a correction for leaked photons which might escape the galaxy from the star-forming regions (for instance, after escaping and contributing to the diffuse ionised gas).

            \item The assumption of a normal IMF in the conversion between the flux and the SFR in certain areas in the collision (like the IC~2163 eyelids as mentioned above) might not be completely adequate. That of a constant SFR is certainly unrealistic as well in the case of interacting galaxies (see the simulation below).
            
        \end{enumerate}

\section{Kinematics}
\label{section:Kinematics}

\subsection{Velocity map}
\begin{figure*}
            \centering
            \includegraphics[width=\textwidth]{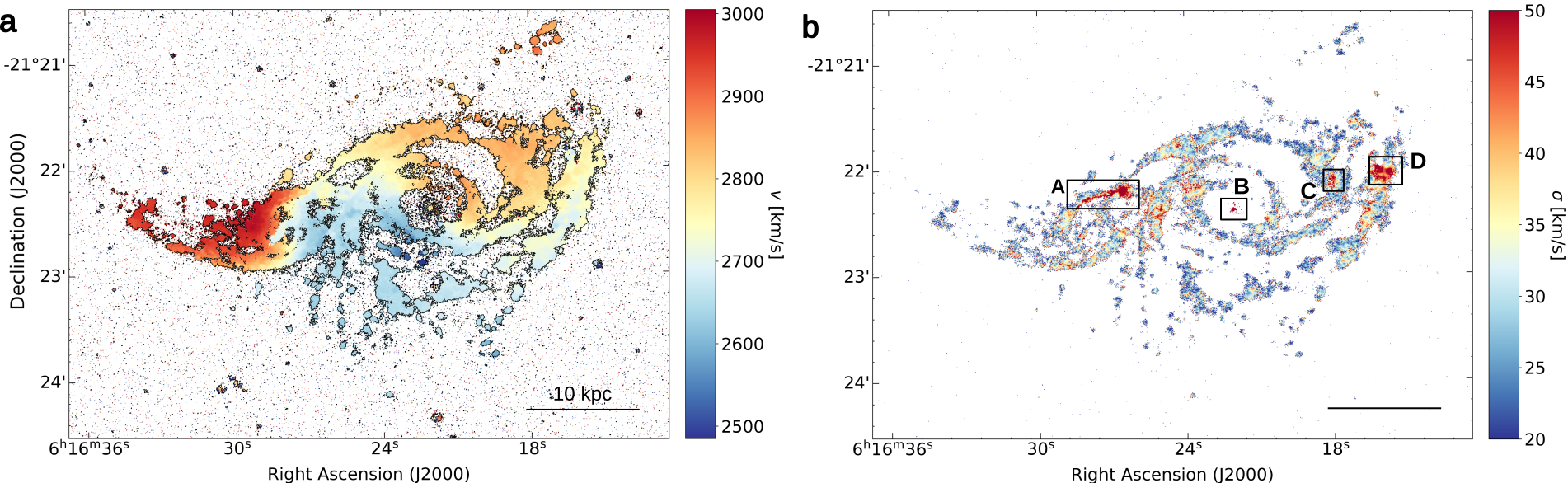}
            \caption{Maps of (a)~the heliocentric velocity, with a black contour corresponding to an H$\alpha$ flux level of $8~\times~10^{-18}$~erg~s$^{-1}$~cm$^{-2}$ and (b)~velocity dispersion, both derived from SITELLE observations. The velocity dispersion map highlights four regions of interest (labeled A to D) characterized by enhanced $\sigma$ values, which are examined in detail in Figure~\ref{fig:VelocityDispersion_boxes}.}
            \label{fig:Maps_sig-vel}
        \end{figure*}

    Figure~\ref{fig:Maps_sig-vel}a presents the heliocentric velocity map of the ionised gas. The large scale velocity distribution can be compared to similar maps derived from \ion{H}{I} and CO observations. In NGC 2207, a warp was reported in the extended \ion{H}{i} disk by \cite{elmegreen1995interactionI}. Our velocity map does not contain enough information in the outskirts of the galaxy to be able to see the warp. Within the inner parts of the disc in NGC 2207, we can infer some asymmetry in the velocity field. On the other hand, the velocity field of the ionised gas in IC~2163 appears disturbed, in particular in the central region and eyelids. This is explained by the large radial and azimuthal streaming motions inferred during the collision, studied in detail by \cite{elmegreen1995interactionI} and \cite{kaufman-ocular} in the case of this system. A detailed modelisation of the velocity structure of the system is beyond the scope of this paper. We have, however, used this map to guide us in the numerical simulations presented in Section~\ref{section:NumericalSimulations}.
    
\subsection{Velocity dispersion}
\begin{figure*}
            \centering
            \includegraphics[width=\textwidth]{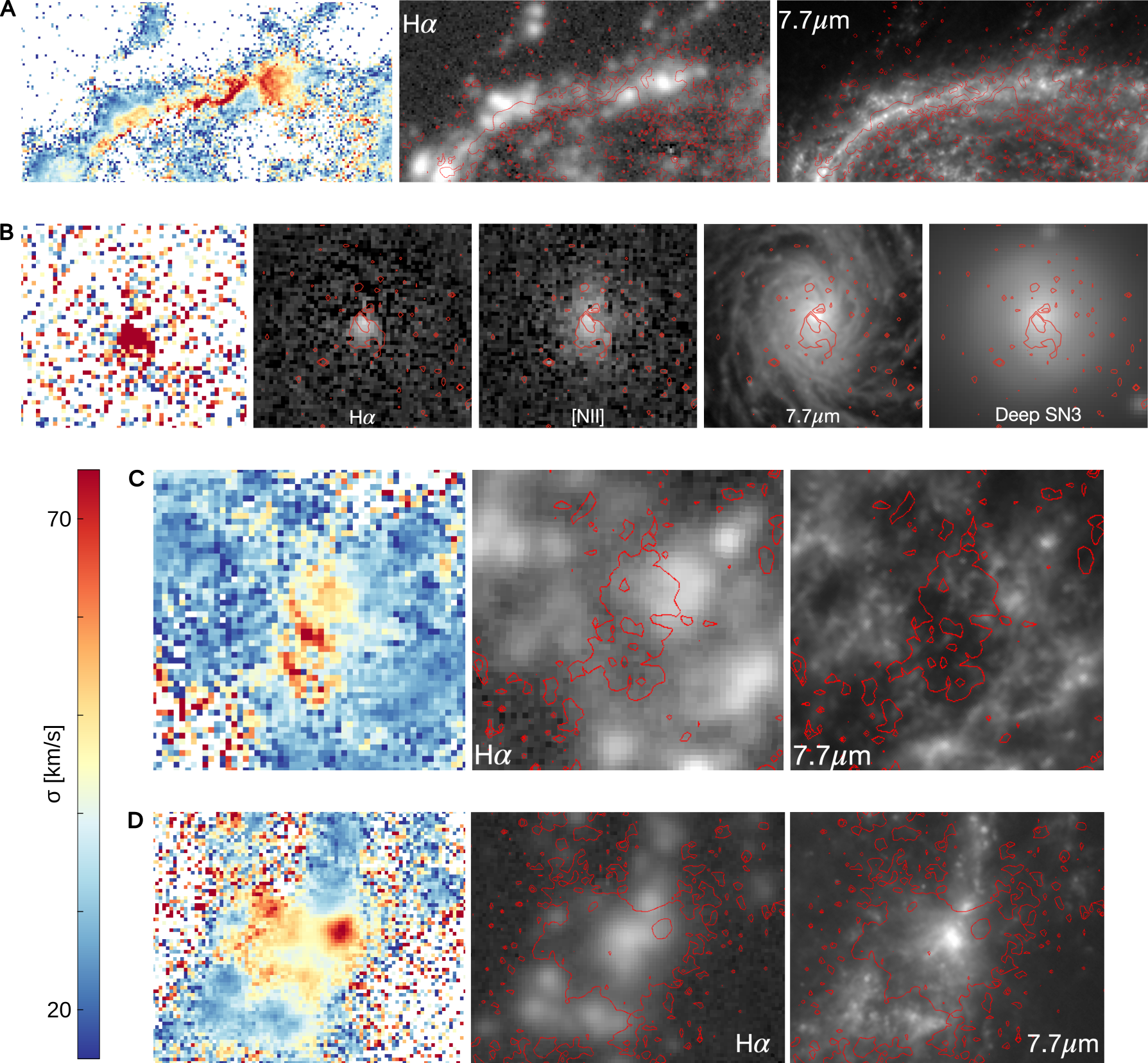}
            \caption{Detailed maps of the four high-$\sigma$ regions identified in Figure~\ref{fig:Maps_sig-vel}b: (A)~the northern eyelid ($53'' \times 26''$), (B)~the nucleus of NGC~2207 ($18'' \times 18''$), (C)~an example of diffuse gas between \ion{H}{ii} regions ($17'' \times 17''$), and (D)~\textit{Feature~i} ($28'' \times 22''$). For regions A, C and D, we present the SITELLE velocity dispersion map with contours at 40 and 65~km~s$^{-1}$, the corresponding H$\alpha$ flux map and the JWST/MIRI F770W image. Panel B additionally includes the [\ion{N}{ii}]6583 flux map and SN3 deep frame, with dispersion contours at 65 and 110~km~s$^{-1}$. The same velocity dispersion color scale is used across all panels.}
            \label{fig:VelocityDispersion_boxes}
        \end{figure*}

    Figure~\ref{fig:Maps_sig-vel}b shows a map of the velocity dispersion, $\sigma$. As expected, $\sigma$ is in the range of $\sim 15$ - $30$ km~s$^{-1}$ in most of the ionised regions, with some noticeable exceptions. We highlight four of them. Figure~\ref{fig:VelocityDispersion_boxes} provides a zoom on the velocity dispersion map along with H$\alpha$, dust (7.7$\mu$m from JWST/MIRI), [\ion{N}{ii}], and continuum images of the nuclear region.

    \textit{Region A} describes an elongated zone near the northern eyelid of IC~2163 where the velocity dispersion is large at $\sim$65~km~s$^{-1}$. \ion{H}{i} kinematics in the area has been studied in detail by \cite{elmegreen1995interactionI}. Their observations indicate strong streaming motions $\sim$100 km~s$^{-1}$ along the eyelid, generated during collision. A mixture of radial and azimuthal flows was also observed in CO in that region \citep{kaufman-ocular}. These streaming motions likely generate large-scale turbulence in the ISM, resulting in the large velocity dispersion seen in our data. Large values in the velocity dispersion in CO in the eyelids were also reported by \cite{kaufman-ocular}. This high level of motion might be responsible for the destruction of older molecular complexes in the eyelids as suggested by \cite{elmegreen2017alma}.
    
\begin{figure}
            \centering
            \includegraphics[width=\columnwidth]{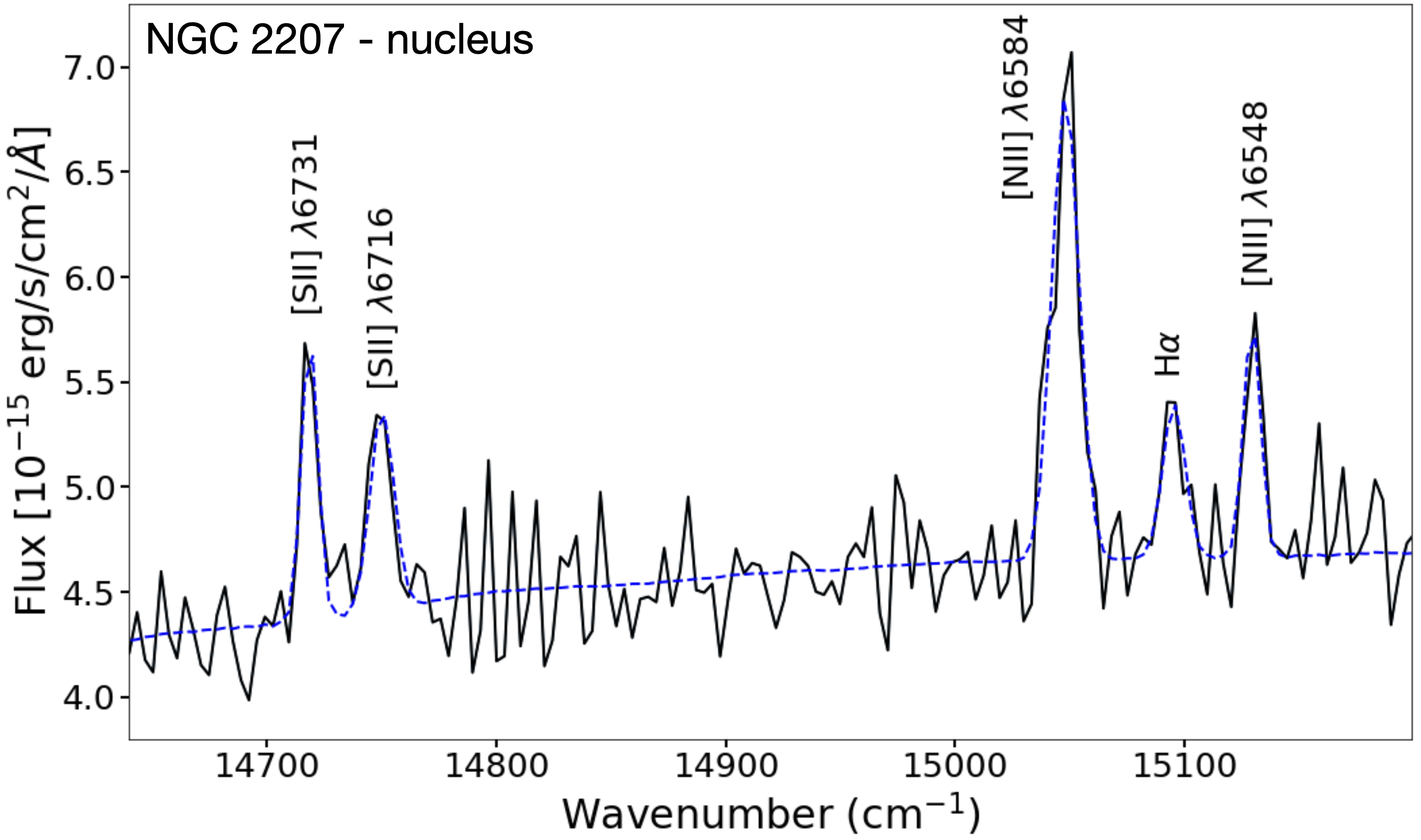}
            \caption{Spectrum of NGC~2207's inner core (radius = 2\arcsec, centered at 06h16m22.03s, $-21^o22\arcmin22\arcsec$) in the SN3 filter. The blue dashed line is the fit with ORCS.}
            \label{fig:NGC2207AGN}
        \end{figure}

    {\it Region B} is the nucleus of NGC~2207. Although it is known to harbor an AGN, mostly because of its X-ray properties, a visible spectrum of this region has never been published, to our knowledge. The properties of its emission lines vary widely spatially: the [\ion{N}{ii}] emission displays a bipolar shape with a North-South orientation ($\sim 4\arcsec\times 7\arcsec$, or $700 \times 1200$~pc); its velocity dispersion is much lower ($\sim 35$ km~s$^{-1}$) in the northern lobe than in the very core and the southern lobe ($\sim 100$ - $150$ km~s$^{-1}$). H$\alpha$ is seen in emission only in the core and the southern lobe, as absorption dominates the northern lobe. We present in Figure~\ref{fig:NGC2207AGN} the SN3 spectrum of the inner 2\arcsec (700~pc in diameter) of the nucleus, which displays clear AGN features: $\sigma \simeq 120$~km~s$^{-1}$, [\ion{N}{ii}]/H$\alpha = 3.7$, [\ion{S}{ii}]/H$\alpha = 2.9$. [\ion{O}{ii}]$\lambda$3727 is also very strong, in particular in the core and the southern lobe. Our SN2 spectrum is much noisier, but [\ion{O}{iii}]$\lambda$5007 is detected while H$\beta$ is only detected in absorption.
    
    {\it Region C} is representative of many similar (mostly DIG) regions across the galaxy, outside of \ion{H}{ii} regions and also corresponding to holes in the distribution of dust as seen in the JWST/MIRI images, where $\sigma$ reaches 50 - 80 km~s$^{-1}$. These could highlight the signs of feedback by supernovae and stellar winds in the recent past.
    
\begin{figure}
            \centering
            \includegraphics[width=\columnwidth]{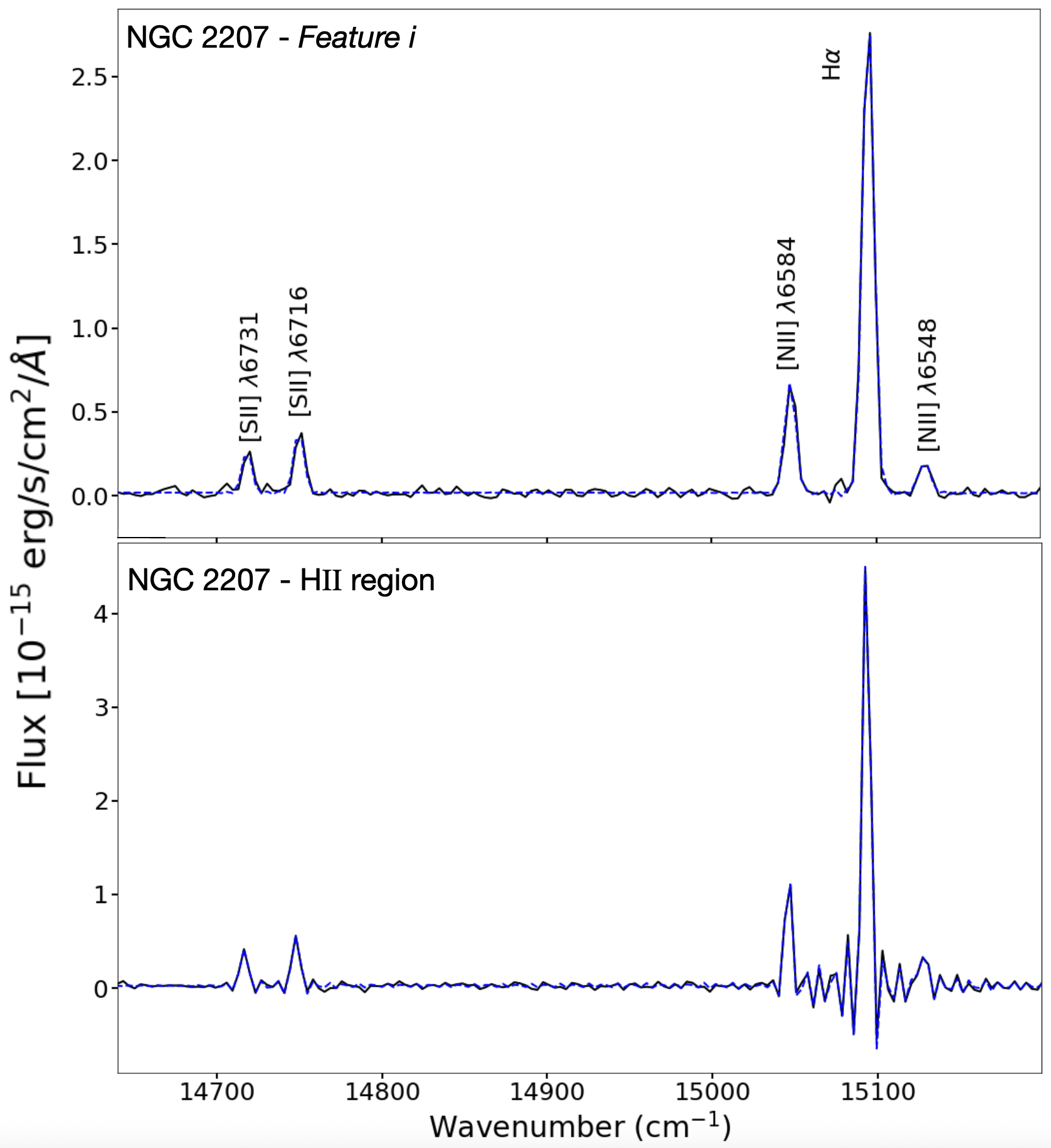}
            \caption{(Upper panel) Spectrum of the high velocity dispersion ($\sigma = 70$~km~s$^{-1}$) region $3.5\arcsec$ west of the core of {\it Feature~i} in NGC~2207 from the SN3 data cube. The blue dashed line is the fit with ORCS. (Lower panel) For comparison purposes, spectrum of an \ion{H}{ii} region close to {\it Feature~i}, with $\sigma = 25$~km~s$^{-1}$. Note the side lobes from the instrumental profile (a sinc function) at the base of the lines, in particular H$\alpha$. These are not visible in the spectrum of the high velocity dispersion region because the line profile is dominated by the gaussian Doppler widening.}
            \label{fig:NGC2207Fi}
        \end{figure}

    Finally, \textit{Region D} is the brightest \ion{H}{ii} region of the galaxy pair, dubbed {\it Feature~i}, which is also the brightest radio and infrared source \citep{kaufman2012ngc} in NGC~2207. The largest velocity dispersion zone in this complex, with $\sigma = 75$~km~s$^{-1}$, is located 3.5\arcsec (600~pc) to the west of \textit{Feature~i}'s core, right above the opaque dust cone discussed by \cite{kaufman20-outflow}; its spectrum is shown in Figure~\ref{fig:NGC2207Fi}, along with that of a nearby \ion{H}{ii} region for comparison. High values of velocity dispersion in {\it Feature~i} are also observed in the CO line \citep{kaufman20-outflow}, but not at the same location. The origin of the increased $\sigma$ in the nebular gas is unclear: champagne outflow following the formation of the young cluster in the core, or increased turbulence caused by stellar winds?

\section{Numerical simulations}
\label{section:NumericalSimulations}

    When simulating a system of two interacting disc galaxies, the parameter space to survey is large. In addition to the intrinsic properties of the galaxies (masses of the various components, radius, scale heights and scale lengths, initial abundances, $\ldots$), there are also the characteristics of the orbit (eccentricity and pericentre), and the initial orientation of the discs relative to the orbit. Fortunately, we can build on the early work of \citet{struck2005grazing}. These authors provide a good estimate of the masses of the galaxies and the characteristics of the orbit, and these are unlikely the change much once additional subgrid physics is added to the algorithm, because gravity is still the main interaction driving the evolution of the system.

    \subsection{Numerical algorithm}
    \label{section:NumericalAlgorithm}
        The collision between NGC~2207 and IC~2163 was modeled using the numerical algorithm GCD+ \citep{kawata2003gcd+, kawata2013calibrating, kawata2014numerical}. GCD+ is a three-dimensional N-body/smoothed particle hydrodynamics (SPH) code which simulates the chemodynamical evolution of galaxies. It incorporates various physical processes including self-gravity, hydrodynamics, radiative cooling, star formation, supernovae feedback, metal enrichment, and metal diffusion. SFR and supernovae feedback are governed by four primary parameters \citep{rahimi2012towards}: the supernova energy output $E_\text{SN} = 1\times10^{51}$~erg (with $10\,\%$ contributing to feedback and the rest dissipated as radiation), the stellar wind energy output $E_\text{SW} = 1\times10^{36}$~erg~s$^{-1}$, the SF efficiency $C* = 0.02$, and the SF density threshold $n_\text{th} = 0.3$~cm$^{-3}$. SF occurs through the conversion of gas particle into star particles, as described in \cite{kawata2014numerical}. These star particles are modeled to represent stars whose masses follow the \cite{salpeter1955luminosity} initial mass function, with their associated metal enrichment from Type Ia and II supernovae calculated according to \cite{woosley1995evolution} and \cite{iwamoto1999nucleosynthesis}.
        
        We point out two important differences between this algorithm and the ones used by \citet{struck2005grazing}. First, there is the treatment of the dark matter haloes. While these authors use static haloes combined with an analytic treatment of dynamical friction, we use fully dynamical haloes represented by dark matter particles. Second, our algorithm includes chemical enrichment; it tracks the abundances of nine chemical elements (H, He, C, N, O, Ne, Mg, Si, Fe) inside the gas and stellar components, and their evolution.

    \subsection{Initial conditions}
    \label{section:InitialConditions}
        For our simulations, a $\Lambda$CDM standard cosmology with $h~=~0.73$, $\Omega_0 = 0.266$, and $\Omega_b = 0.044$ is assumed. Each modeled galaxy consists of a dark matter halo represented by dark matter particles, along with two galactic disks, each made up of gas and star particles. The dark matter halo is described by a Navarro-,Frenk-White (NFW) profile \citep{navarro1996structure} with a concentration parameter $c = 20$. The galactic disks are configured according to an exponential surface density profile 
            \begin{equation}
                \rho = \frac{M}{4\pi\zeta l^2} \,\mathrm{sech}^2\left(\frac{z}{\zeta}\right)\exp{\left(-\frac{R}{\ell}\right)} \;,
            \end{equation}
    
        \noindent where $M$ is the gas/star disk mass, $\ell$ is the scale length, $\zeta$ is the scale height (set to $l/10$), and $R$ and $z$ correspond to the radial and vertical coordinates, respectively. The masses, scale lengths, and scale heights of the various components, for the best model, are listed in Table~\ref{tab:initial1}. The initial radial metallicity profile for both the stellar and gaseous population within the two galaxies is set by the iron abundance. The primary galaxy (namely \textit{Galaxy A}) has an initial central iron abundance of $\rm[Fe/H]=-0.09$, while the satellite galaxy (\textit{Galaxy B}) has [Fe/H] = 0.1, with corresponding radial gradients of $-0.03$~kpc$^{-1}$ and $-0.03$~kpc$^{-1}$, respectively. Additionally, $\alpha$-elements are present solely in the stellar component, with abundances expressed by
            \begin{equation}
                [\alpha/\text{Fe}] = -0.16[\text{Fe}/\text{H}] \;.
            \end{equation}
        
        Galaxy~B starts with a lower stellar mass and higher metallicity than Galaxy~A, hence they do not follow the mass-metallicity ($M_*$-$Z$) relation \citep{tremontietal2004}. Note, however, that a difference of 0.19 in the central values of \FeHratio\ is comparable to the scatter in the $M_*$-$Z$ relation. Several scenarios could explain how Galaxy~B had a high metallicity before interacting with Galaxy~A. For instance, if it had experienced an even earlier encounter with a massive galaxy, tides could have removed stars and low-metallicity gas located in the outskirts of Galaxy~B, thus reducing $M_*$ and increasing $Z$ (see \citealt{Williamsonetal2016}).
        
\begin{figure}
                \centering
                \includegraphics[width=\columnwidth]{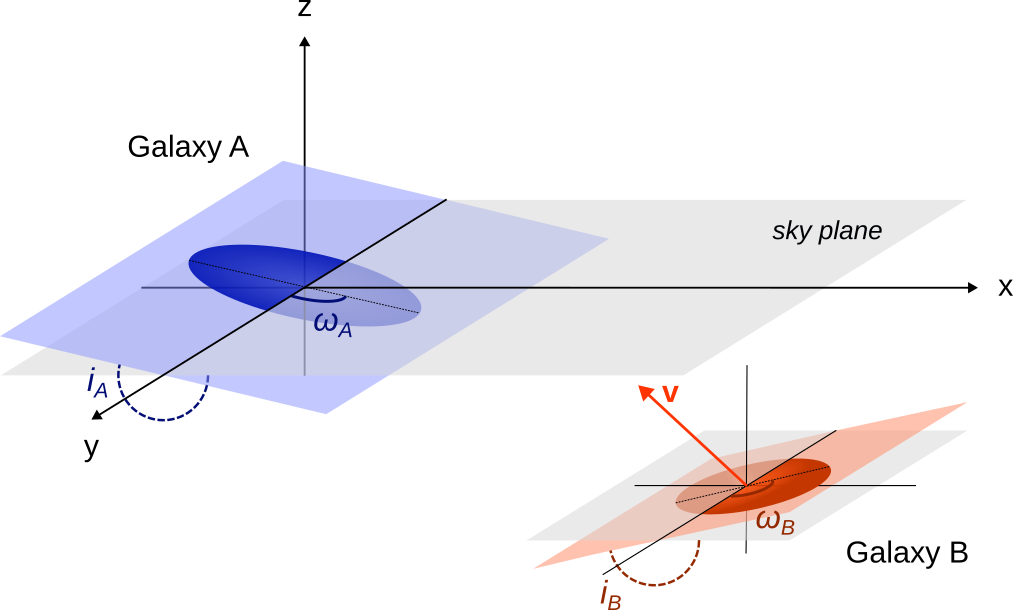}
                \caption{Initial configuration of the system. Galaxy A is at rest, while Galaxy B is moving with a velocity $\mathbf{v} = (v_x, v_y, v_z)$. The inclination angles $i$ and $\omega$ correspond to rotations of the plane around the $y$-axis and $z$-axis, respectively. For instance, an angle $i$ around the $y$-axis indicates a clockwise rotation of the plane in the $x$-$z$ plane.}
                \label{fig:GeometrySimulation}
            \end{figure}

        The initial configuration of the system is illustrated in Figure~\ref{fig:GeometrySimulation}. For the orbital parameters, we started with the initial positions and velocities obtained from the best model presented by \cite{struck2005grazing}, which reproduces several key features of the NGC~2207 and IC~2163 observations.
        Subsequently, several dozen simulations were then conducted to first adjust the initial conditions for the algorithm used. Among other factors, the initial distance between the galaxies must be sufficient to allow a compromise between allowing for a nearly isolated stage and interaction before the formation of substructures (e.g., bars) in their disks, while also ensuring that this distance is not too large as the spiral arms may eventually evolve into central stellar bars within $\sim1$~Gyr \citep{fanali2015bar}. Additional adjustments to the initial conditions were made to meet our objectives, specifically: to reproduce the overall morphology, kinematics, metallicity distribution, and global SFR. The selected best model is the one that most accurately reproduces the morphological and kinematic features of the observations, chosen from a set of few dozen models with different initial conditions. The initial position and velocities of the galaxies, and the orientation of their discs, are listed in Table~\ref{tab:initial2}. In our simulations, Galaxy A and Galaxy B correspond to the interacting pair NGC~2207 and IC~2163, respectively. Galaxy A is fixed at the origin throughout the entire simulation. The entire model was rotated by $85^\circ$ around the $z$-axis after the run (i.e., $85\degr$ clockwise in the $x$-$y$ plane) to approximately align with the orientation of the observations. Here, the $x$-$y$ plane is equivalent to the sky plane.
        
        To investigate the specific effects of the interaction, we performed additional simulations in which each galaxy was evolved in isolation, using the same initial conditions as in the interacting run. These isolated simulation, referred to as \textit{Galaxy A isolated} and \textit{Galaxy B isolated}, were carried out with a factor of three fewer particles in each disk (halo, gas, stars) compared to the high-resolution interacting simulation. This lower resolution was chosen to reduce computational cost, as a higher particle number was unnecessary. Table~\ref{tab:resolution} gives, for each galaxy and each resolution used, the number of particles, $N$, and the mass per particle $m$, which determines the mass resolution of the simulations.
        
            \begin{table*}
                \caption{Initial properties of the simulated galaxies. $M$, $\ell$, and $\zeta$ are the mass, disc scale length and scale height (component indicated as index) respectively. Masses are in units of ${\rm 10^{10}M_\odot}$. Scales are in units of $\rm kpc$.} \label{tab:initial1}
                \centering
                \small
                \begin{tabular}{llcccccccc}
                \hline
                \smallskip
                Galaxy & Representing & $M_{\rm halo}$ & $M_{\rm gas}$ & $M_{\rm star}$ & $\ell_{\rm gas}$& $\zeta_{\rm gas}$  & $\ell_{\rm star}$ & $\zeta_{\rm star}$ \\
                \hline
                Galaxy A & $\rm NGC\,2207$ & 387 & 0.875 & 12.63 & 5.21 & 0.521 & 1.125 & 0.1125 \\
                Galaxy B & $\rm IC\,2163$  & 194 & 0.249 & 5.751 & 3.60 & 0.36  & 0.6   & 0.06 \\
                \hline
                \end{tabular}
            \end{table*}
            
            \begin{table}
                \caption{Initial positions $\bf R$ and velocities $\bf V$ of the galaxies, and orientation of their disks. Indices A and B refer to Galaxy A and Galaxy B, respectively. $\bf R$ and $\bf V$ are expressed in cartesian coordinates $(x,y,z)$, while angles $\theta$ and $\phi$ indicate, in spherical coordinates, the direction of the angular momentum vector of the disks.}\label{tab:initial2}
                \centering
                \small
                \begin{tabular}{lr}
                \hline
                ${\bf R}_{\rm A}\, [{\rm kpc}]$ & (0,0,0)\\
                ${\bf R}_{\rm B}\, [{\rm kpc}]$ & ($85,-25,4$)\\
                ${\bf V}_{\rm A}\, [\rm km\,s^{-1}$] & (0,0,0)\\
                ${\bf V}_{\rm B}\, [\rm km\,s^{-1}$] & ($-25,185,4$)\\
                $(i,\omega)_{\rm A}$ & $(195^\circ,15^\circ)$\\
                $(i,\omega)_{\rm B}$ & $(305^\circ,230^\circ)$\\
                \hline
                \end{tabular}
            \end{table}
            
            \begin{table}
                \caption{Number of particles and particle masses for the low-resolution runs (LR) and the high-resolution run (HR). Gas and star particles have the same mass $m_{\rm b}$ (the baryon mass). Masses are in units of ${\rm 10^3M_\odot}$}\label{tab:resolution}
                \centering
                \small
                \begin{tabular}{lcccccc}
                \hline
                \smallskip
                Galaxy & $N_{\rm halo}$ & $N_{\rm gas}$ & $N_{\rm star}$ 
                & $m_{\rm halo}^{\phantom0}$ & $m_{\rm b}^{\phantom0}$ \\
                \hline
                Galaxy A (LR) & $197\,590$ & $8\,750$  & $126\,250$  & $19\,587$ & $1\,000$ \\
                Galaxy B (LR) & $98\,795$  & $2\,490$  & $57\,510$   & $19\,839$ & $1\,000$ \\
                Galaxy A (HR) & $592\,770$ & $26\,250$ & $378\,750$  & $6\,529$  & 333 \\
                Galaxy B (HR) & $296\,385$ & $7\,470$  & $172\,530$  & $6\,613$  & 333 \\
                \hline
                \end{tabular}
            \end{table}

    \subsection{Simulations}
    \label{section:Simulations}
        \subsubsection{General Evolution}
        \label{section:Simulations_GeneralEvolution}
\begin{figure*}
                    \centering
                    \includegraphics[width=0.85\textwidth]{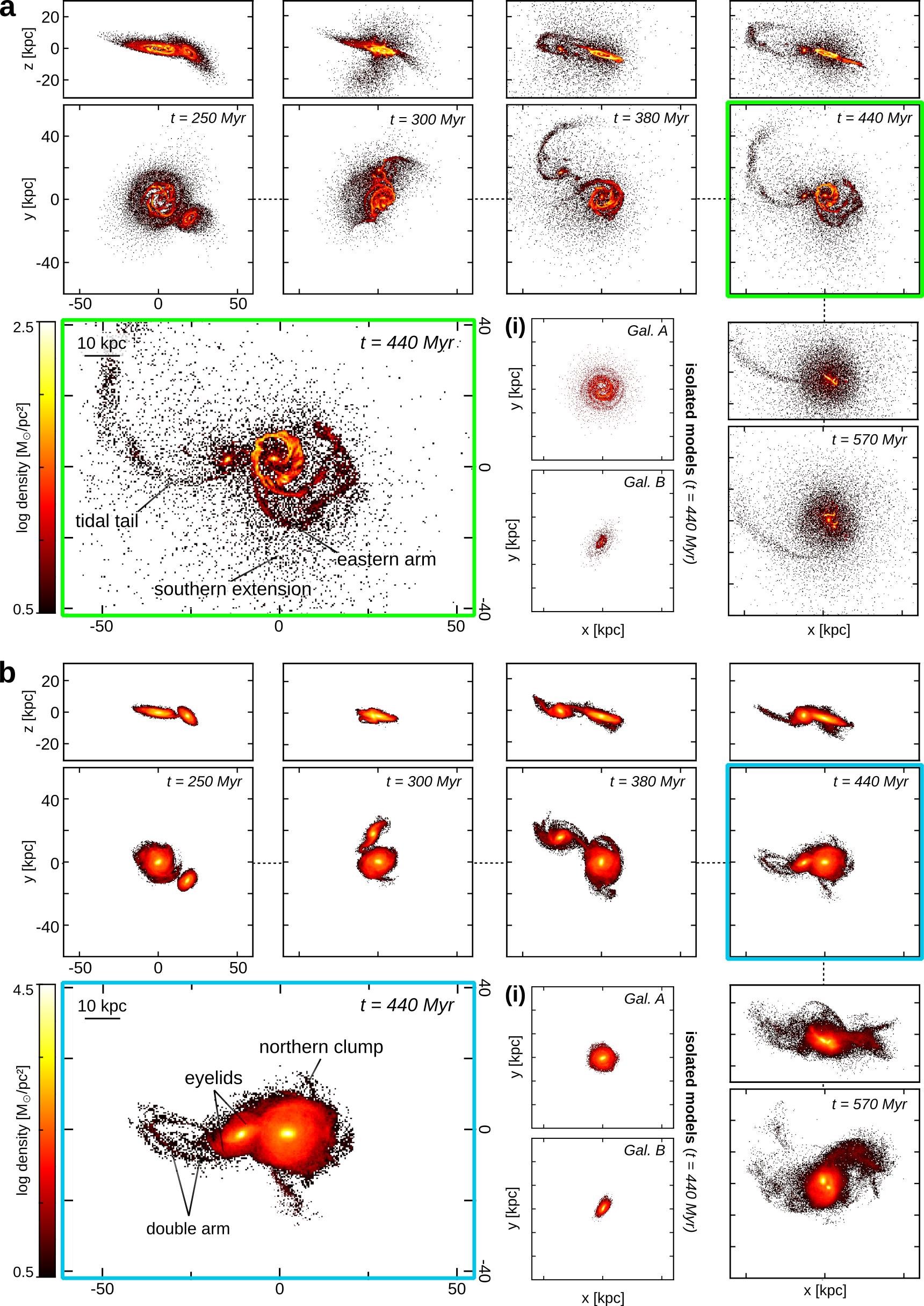}
                    \caption{Temporal evolution of the best model for the (a)~gas and (b)~stellar disks, projected onto the $x$-$y$ (sky plane) and $x$-$z$ planes and color-coded by density. The ``present'' time is highlighted within the temporal sequence (connected by dots) and magnified in the respective lower-left corner with key morphological structures identified (sky plane). (i) $x$-$y$ plane projections of the two galaxies for isolated simulations at the ``present'' time.}
                    \label{fig:BestSimulation}
                \end{figure*}

            Figure~\ref{fig:BestSimulation} presents the temporal evolution of the best model for the star and gas particles in the disks projected onto the $x$-$y$ and $x$-$z$ planes. The simulation begins when the centre of Galaxy~B is located at a distance of approximately $89$~kpc from the centre of Galaxy~A with a relative velocity of $185$~km~s$^{-1}$ in the $x$-$y$ plane. The first contact occurs around $280$~Myr, as the gas disk of Galaxy~B grazes that of Galaxy~A in a counterclockwise motion, moving from the southwest to the north. Following their initial interaction, Galaxy~B moves away to a maximum distance of $\sim32$~kpc, while a prominent tidal arm remains connected to Galaxy A. It later approaches Galaxy~A again, this time from the eastern side. At $t = 440$~Myr, the relative position and morphological features of the two galaxies exhibit the greatest similarity to the observations. Comparing the large panels with green and blue edges in Figure~\ref{fig:BestSimulation} with the image shown in Figure~\ref{fig:NGC 2207-FOV}, we see that the tidal tail, eyelids, spiral arms, southern extension, and northern clump are all well-reproduced. We therefore define this moment in the simulation as the \textit{``present'' time}. 
    
            The overall morphology is consistent with the one obtained by \citet[see their Figure~10, right panels]{struck2005grazing}. This was expected, since were are using similar initial masses and orbital parameters. In addition, we reproduce structures found in the observations, that were not reproduced in their simulations, namely the double arm and the northern clump. We note that these authors identified the present time as $t=380\,\rm Myr$ while ours is $440\,\rm Myr$. We use the same orbital parameters, but start the simulation at an earlier stage, when the initial separation between the galaxies is larger. While this is the main reason for this difference in ``present'' times, using dynamical dark matter haloes, with a full N-body treatment of gravity might also have some effect on the timing of the interaction.

        \subsubsection{Star formation}
        \label{section:Simulations_SF}
\begin{figure}
                    \centering
                    \includegraphics[width=\columnwidth]{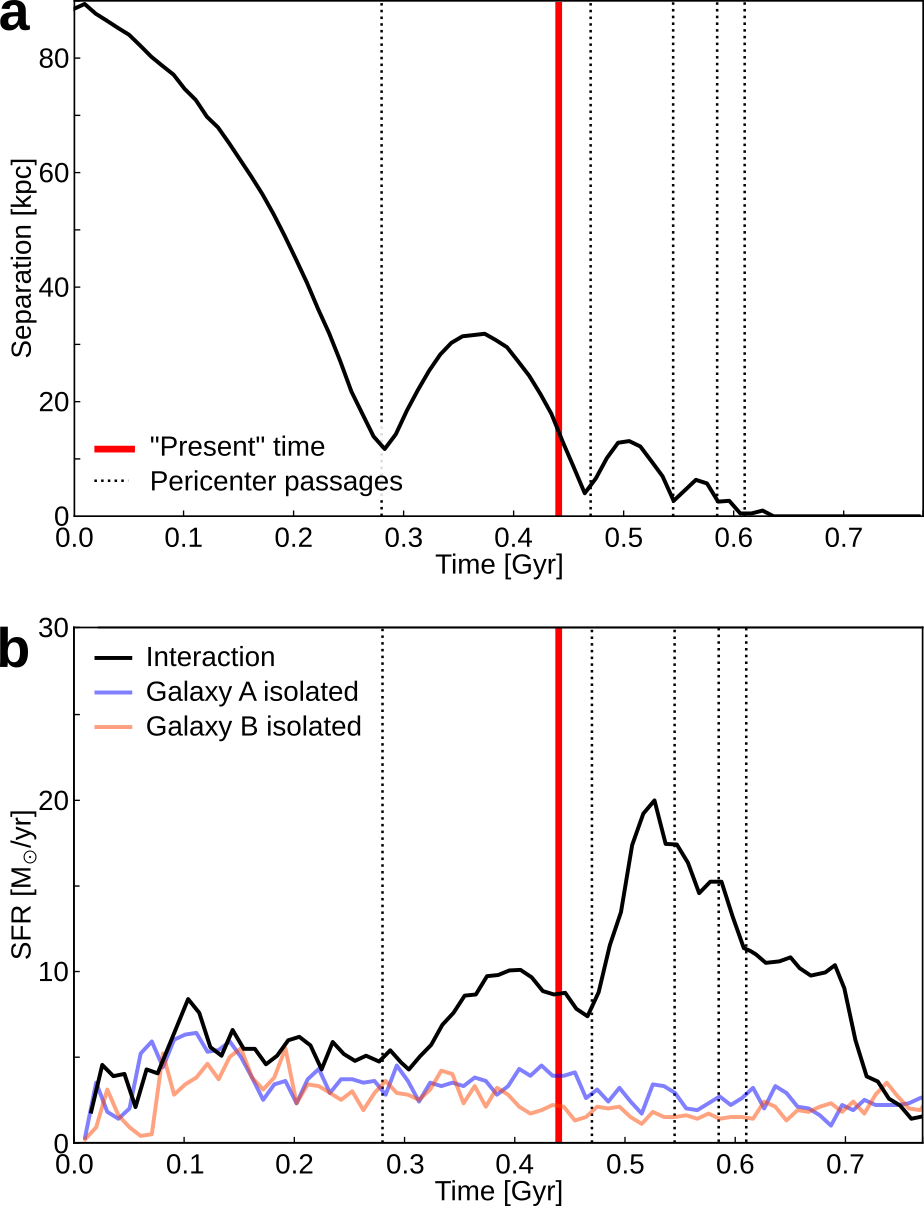}
                    \caption{Temporal evolution of the separation between the interacting galaxies in the best model, measured in the $x$-$y$ plane. (b)~Temporal evolution of the SFR for the interacting system (black; total for both galaxies), compared to the isolated evolution of Galaxy A (blue) and Galaxy B (orange). In both panels, black vertical dotted lines indicate the times of pericentric passages, while the thick red vertical line marks the ``present'' time ($t=440\,\rm Myr$).}
                    \label{fig:Sep-SFR-Simulation}
                \end{figure}

            Figure~\ref{fig:Sep-SFR-Simulation}a shows the evolution of the projected separation between the two galaxies in the $x$-$y$ plane throughout the simulation\footnote{This is essentially the same as the 3D separation, as the $z$-separation remains relatively small.}. The SF history in our simulations is tracked by computing the SFR between successive snapshots. The temporal evolution of the SFR is shown in Figure~\ref{fig:Sep-SFR-Simulation}b for both the interacting system -- where SFRs from the two galaxies are summed -- and the galaxies simulated in isolation. In both panels, the dotted lines identify the pericentre passages, when the galaxies are the closest. Prior to the first pericentre passage (i.e., before $\sim280\,\rm Myr$), all simulations exhibit similar SFRs, typically around $\rm5\,M_\odot\,yr^{-1}$. Following this first encounter, the interacting case diverges from the isolated ones: while the isolated galaxies experience a gradual decline in SF activity, the interacting system undergoes a sequence of starburst episodes, each occurring shortly after a pericentre passage. At each close passage, the tidal field each galaxy exerts on the other one compresses the gas, triggering a starburst \citep{renaudetal2014}.The first burst reaches slightly $\rm10\,M_\odot\,yr^{-1}$. The ``present'' time occurs just after the peak of this burst, with a global SFR of $\rm\sim9\,M_\odot\,yr^{-1}$, halfway between the values determined from the observations of the ionised gas and the dust emission (Section~\ref{section:SFRs}). Following this, we observe a second and more prominent enhancement in the SFR after the subsequent pericentre passage, reaching a peak of 
            $\rm20\,M_\odot\,yr^{-1}$ at $\rm\sim530\,Myr$. At this stage, the two galaxies are separated by approximately 10~kpc. This peak value is about four times higher than the sum of the SFRs of the two galaxies when evolved in isolation. After this maximum, the SFR decreases approximately linearly as the galaxies continue their mutual orbital motion, gradually losing separation and angular momentum, until coalescence at $\rm\sim630\,Myr$, when the individual galactic centers can no longer be distinguished. The decrease in SFR after $\rm530\,Myr$ is caused by the depletion of the gas supply by the SF process itself, and by gas being pulled out of the galaxies by tides.
        
            We acknowledge that the exact values of the simulated SFRs are sensitive to several parameters. While the SF efficiency and density threshold directly affect the SFR, the initial gas density profile -- set by the gas mass and scale length -- also plays a significant role, as expected. Additionally, resolution effects are present: simulations with higher particle numbers tend to exhibit slightly elevated SFRs, as they better resolve dense star-forming regions. However, these differences remain modest, and the key result in this analysis lies in the overall shape and relative evolution of the SF histories, rather than their absolute values.

        \subsubsection{Metallicity}
        \label{section:Simulations_Metallicity}
            The GCD+ code self-consistently tracks the chemical enrichment of gas and stellar particles, including key elements such as hydrogen, oxygen, and iron. Number abundances for each element were computed by dividing the elemental mass fractions of gas particles by the corresponding atomic mass. This allows us to study the impact of galaxy interaction on the chemical properties of the system. To enable a meaningful comparison with observations, we adjusted the initial radial iron abundance profiles in both galaxies (see Section~\ref{section:InitialConditions}) so that the resulting oxygen abundance distribution matches the observational data derived from SITELLE. Observed abundances were derived using several strong-line calibrations (see Section~\ref{section:OxygenAbundance}); for consistency with the simulations, we adopt the O3N2 calibration of \citetalias{marino2013o3n2} as a reference, without particular preference.  
            
\begin{figure*}
                    \centering
                    \includegraphics[width=\textwidth]{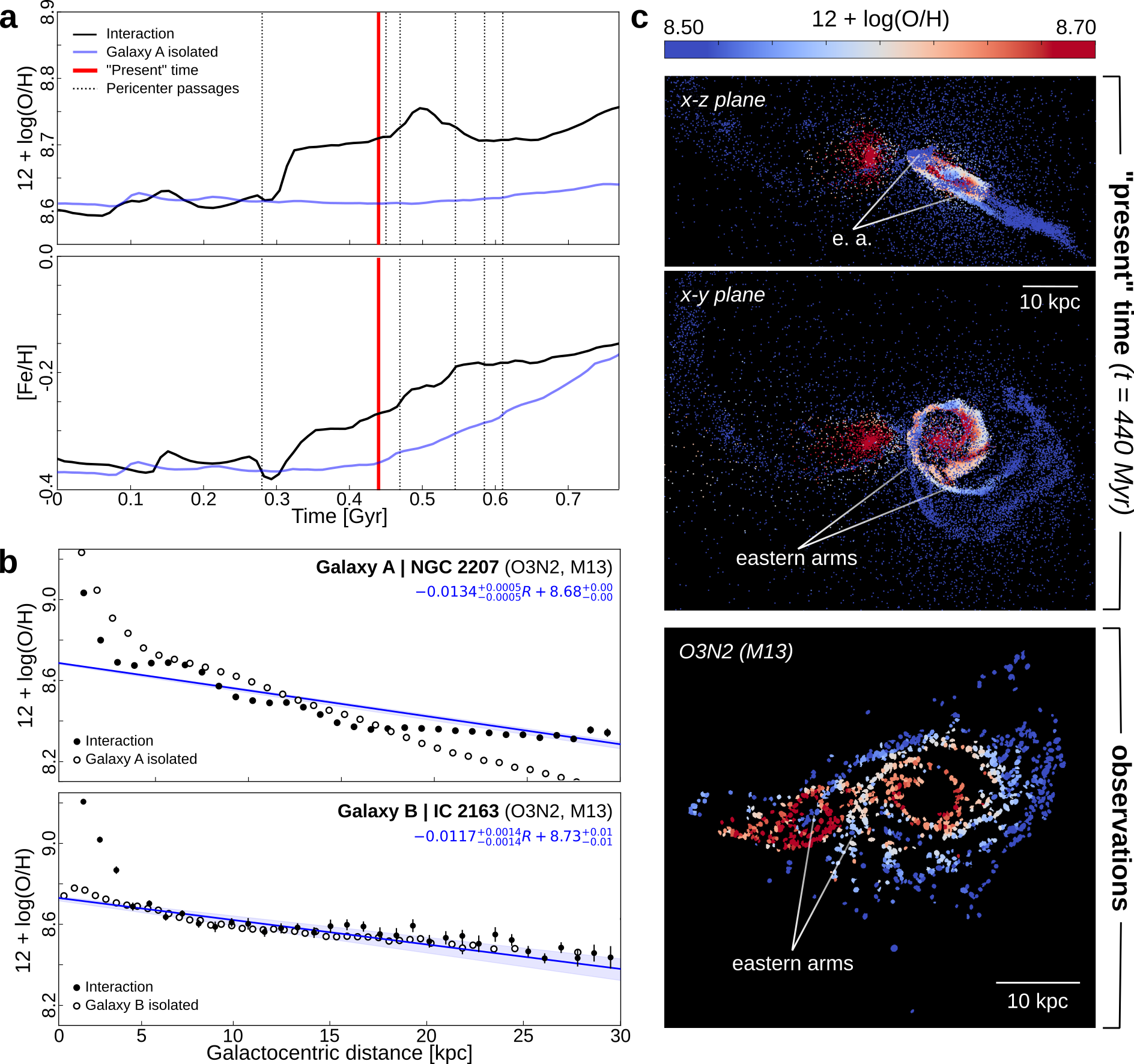}
                    \caption{(a)~Temporal evolution of the median oxygen abundance (top) and iron abundance (bottom) for both simulated galaxies (black) and the Galaxy A isolated (blue). Vertical dotted lines mark pericentric passages, and the thick red line indicates the ``present'' time. (b)~Radial oxygen abundance profiles as a function of galactocentric distance for the individual galaxies, with Galaxy A/NGC 2207 shown in the top panel and Galaxy B/IC~2163 in the bottom panel. Observational data, derived from the O3N2 \citepalias{marino2013o3n2} calibration (see Section~\ref{section:OxygenAbundance} for details), are shown as blue linear regression fits with associated confidence intervals. Black markers represent the mean oxygen abundance measured in each concentric annulus for the interacting simulation, while unfilled markers correspond to the isolated galaxy simulations, both at the ``present'' time. (c)~Maps of \OHratio\ abundance. The two upper panels show the abundance distributions from the interacting simulation in the $x-z$ and $x$-$y$ planes. The lower panel shows the abundance map derived observationally from the O3N2 \citepalias{marino2013o3n2} calibration using integrated fluxes from the final emission domains. All maps share the same color scale.}
                    \label{fig:MetallicitySimulation}
                \end{figure*}

            We first examine the time evolution of the gas-phase metallicity in the simulation. We track the median values of both \OHratio\ and iron metallicity \FeHratio\ across snapshots. As the galaxies become increasingly mixed at later times, global values are computed for the full system. For comparison, we use the isolated evolution of Galaxy A as a control case, given its higher gas content. 
            Figure~\ref{fig:MetallicitySimulation}a shows the time evolution of \OHratio\ and \FeHratio\ for the entire system (black lines) and for Galaxy A when run in isolation (blue lines). In both cases, the oxygen and iron abundances steadily increases with time, as gas get enriched by stellar outflows. In this isolated scenario, both oxygen and iron abundances exhibit a steady and smooth increase over time. In both scenarios, there are some occasional drops, more noticeably in the simulation with interaction. SF and chemical enrichment tends to take place in the same high-density regions. Gas enriched by a first generation of stars can later be removed when a second generation of stars form. Since this does not happen in low density regions, the overall metallicity can drop. At first pericentre passage ($t = 280\,\rm Myr$), the starburst results in a sudden increase in oxygen abundance, and a progressive increase in iron abundance. Alpha elements, including oxygen, are
            produced mostly by core-collapse SNe. Their progenitors have a short lifetimes, which explains the sudden increase in oxygen abundance. Iron is produced mostly by Type~Ia SNe whose progenitors have longer lifetimes, hence iron enrichent is slower and more extended in time. The second starburst, taking place at ($t = 460\,\rm Myr$), is not as strong, but we still observe a sudden increase in oxygen abundance and a progressive increase in iron abundance. The final increases in abundance at $t>0.65\,\rm Gyr$, after the galaxies have merged, is likely caused by low-metallicity gas located in peripheric regions escaping the system.
            
            To compare the ``present'' time radial trends, we extract the \OHratio\ radial profiles for each galaxy. Since the host galaxy of each particle is not explicitly identified in the simulation, we approximate the separation of the two systems spatially by dividing the domain in the $x$–$y$ plane and excluding the central overlapping region.  Radial oxygen abundance profiles were then constructed by averaging the values of \OHratio\ for gas particles within concentric annuli centered on each separated galaxy. These simulated profiles are shown in Figure~\ref{fig:MetallicitySimulation}b, alongside the observational linear fits derived from SITELLE data (see Section~\ref{section:OxygenAbundance} for details). The inner $\rm\sim2.5\,kpc$ are not probed observationally due to the absence of detected \ion{H}{ii} region complexes, but beyond this radius, the interacting simulations broadly reproduce the observed abundance gradients. Small-scale fluctuations appear in the simulated profiles, likely due to azimuthal variations (e.g., intersecting arms or localized features) in the annuli. Comparing the isolated runs with the interacting one, Galaxy A in the interacting run shows a shallower gradient at larger radii (> 15~kpc), consistent with dilution due to gas inflow. For Galaxy B, the gradient is largely similar between isolated and interacting cases, although the interacting case shows slightly higher oxygen abundances at nearly all radii. This enhancement is more pronounced toward the center, likely due to interaction -- triggered SF concentrated in the compact central region (within $\sim$5~kpc), including the eyelids. At larger radii, where the gas density is lower, SF -- and thus chemical enrichment -- is less efficient. 

            Figure~\ref{fig:MetallicitySimulation}c further presents the spatial distribution of \OHratio\ in both simulated and observed galaxies. The overall metallicity patterns are well reproduced, with Galaxy B (IC~2163) showing a steep gradient from a central peak to lower abundances in its tidal tail, and Galaxy A (NGC 2207) displaying a smoother decline toward its outer arms and southern extension. In both the simulation and observations, the eastern arm of Galaxy A crosses -- or lies close in projection to -- the central region of Galaxy B. The simulation confirms that this overlap contributes to the lower oxygen abundance observed near the center of IC~2163, which in fact originates from gas belonging to NGC~2207. This reinforces the idea that metallicity serves as an effective tracer for disentangling such overlapping systems in projection (see Section~\ref{section:Separation}).

        \subsubsection{Kinematics}
        \label{section:Simulations_Kinematics}
\begin{figure*}
                    \centering
                    \includegraphics[width=0.9\textwidth]{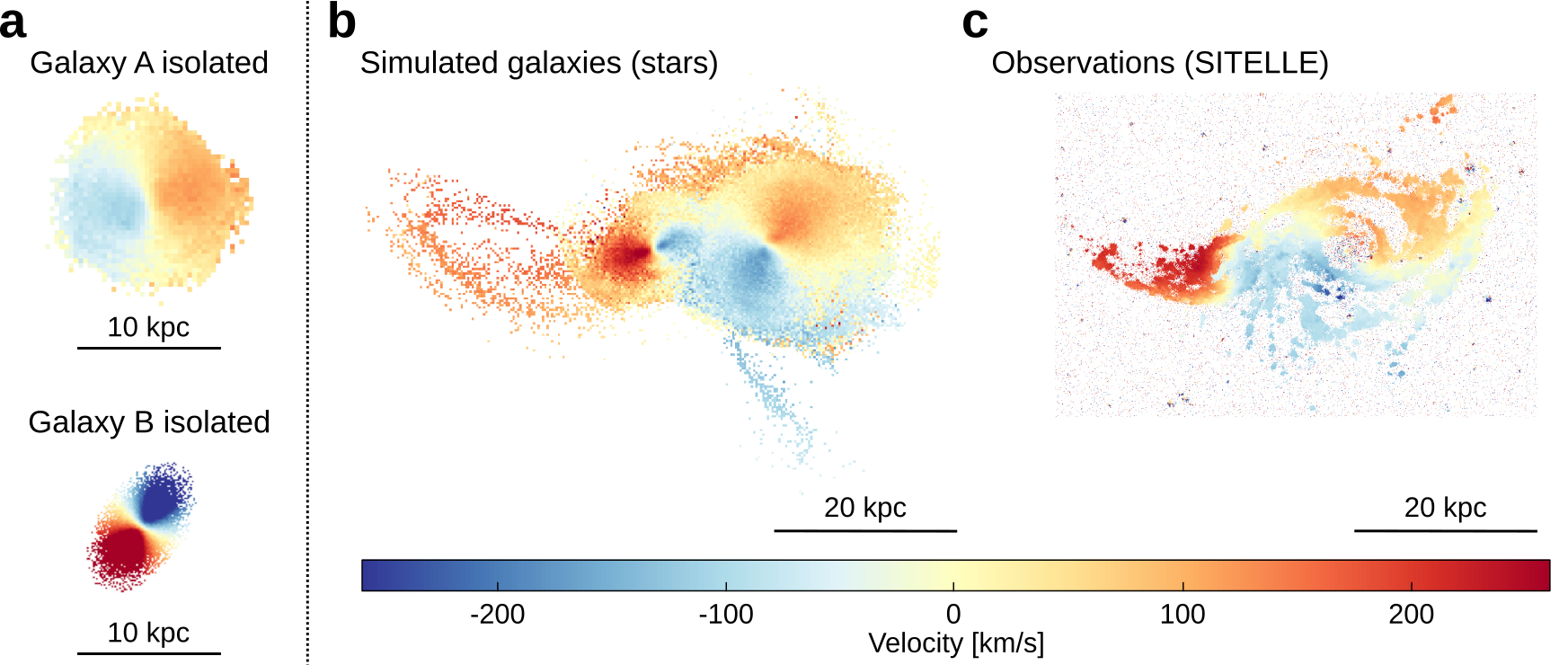}
                    \caption{(a)~LOS velocity fields of the simulated galaxies evolved in isolation, with Galaxy A (top) and Galaxy B (bottom). (b)~LOS velocity map of the interacting system at the ``present'' time snapshot ($t = \SI{440}{Myrs}$). (c)~Observed velocity map from SITELLE, shown relative to a velocity of $2750\,\kms$. The same color scale is applied to all panels. Only the simulated stellar velocity map is shown here, while the gas map is presented in Figure~\ref{fig:KinematicsSimulation-gas} owing to its lower spatial sampling.}
                    \label{fig:KinematicsSimulation}
                \end{figure*}

            To obtain simulated velocity maps comparable to the observations, we primarily adjusted the relative positions and inclination angles ($i$ and $\omega$) of the two galaxies. For each simulation snapshot, we extracted the LOS velocities of star particles (chosen for their higher sampling density; results are similar when using gas particles) along the $z$-axis and projected them onto the sky plane ($x$-$y$). The velocity maps were computed as bi-dimensional histograms, where each spatial bin represent the mean LOS velocity of the particles it contains. Figure~\ref{fig:KinematicsSimulation} show the resulting velocity fields for both galaxies simulated in isolation and during the interaction at the ``present'' time, alongside the SITELLE velocity map for comparison. To match the simulation reference frame, a systemic velocity of $2750\,\kms$ was subtracted from the observed data.        
        
        The interacting simulation reproduces several key kinematic features seen in the SITELLE map. Positive velocities of $\sim200\,\kms$ are recovered in the tidal tail of Galaxy B, while velocities of $-$150 to $-100\,\kms$ are found in the central-southern region of the system. In Galaxy A, the north-western side reaches $\sim100\,\kms$, and a southern arm -- analogous to the eastern arm of NGC~2207 passing over the ocular of IC~2163 -- is reproduced with a similar amplitude. Minor discrepancies remain in the overlap region between the two galaxies, likely due to the averaging procedure applied in the simulated velocity map, which tends to bias intermediate velocities toward zero. 

        A comparison with isolated galaxy simulation  highlights the impact of the interaction on the global kinematics. Initially, the zero-velocity curves (ZVCs) - separating the approaching and receding sides - are oriented at $\sim105\degr$ and $\sim145\degr$ (west through north) for Galaxy A and Galaxy B, respectively, and remain unchanged throughout the isolated runs. In the interacting case, these orientations persist until $\sim$280~Myr, when Galaxy B grazes the western side of Galaxy A and begins an anticlockwise passage to north. After this first pericenter passage, the ZVCs rotate to $\sim135\degr$ for Galaxy A and $\sim105\degr$ for Galaxy B. During the second pericenter passage ($t = 440$~Myr), the velocity fields become highly disturbed, and no clear large-scale rotation patterns are distinguishable.
        
\section{A quadruple system?}
\label{section:quadruple}
    Dozens of (mostly background) galaxies are seen in the SITELLE deep images;  the two brightest ones, identified with arrows in Figure~\ref{fig:NGC 2207-FOV}, are particularly interesting in the context of the interaction studied in this paper.
    
    \subsection{A warped dwarf star-forming galaxy}
    \label{section:SmallGalaxy}
\begin{figure*}
                \centering
                \includegraphics[width=\textwidth]{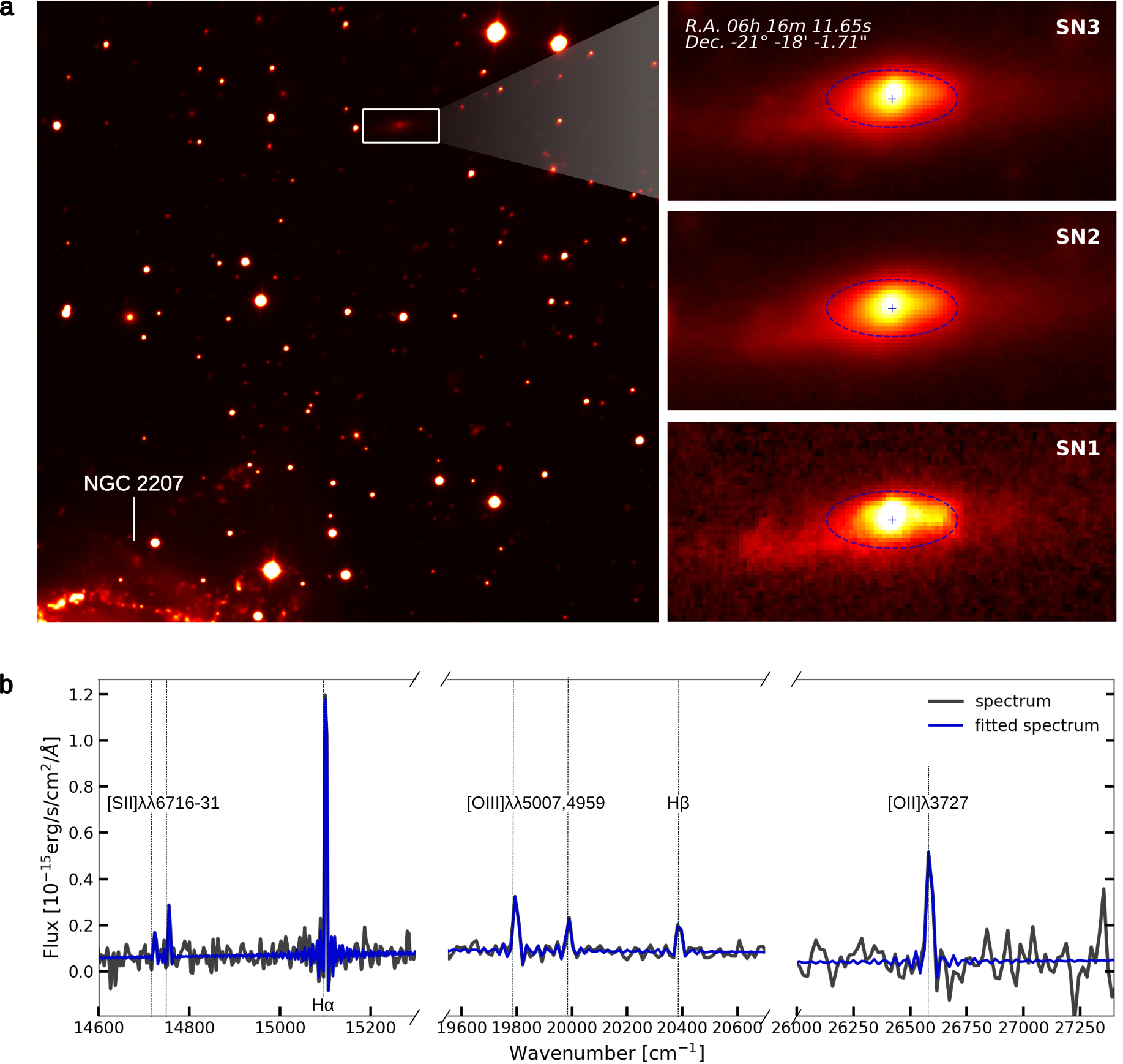}
                \caption{(a)~On the left, the SN3 deep frame provides the location of the object int the northwest region of the FOV in relation to other galaxies. On the right, a close-up of the potential galaxy int the SN3, SN2, and SN1 deep frames with an ellipse marking the emission region from which the spectra were extracted. (b)~Post sky-subtraction spectra from the SN3, SN2, and SN1 filters with the main emission lines identified and fitted using \texttt{ORCS}.}
                \label{fig:MysteriousGalaxy}
            \end{figure*}

        In the northwest part of the SITELLE FOV, at a projected distance of 45 kpc from the center of NGC~2207, stands an elongated galaxy (centered on R.A. 06h16m11.65s, Dec. -21°18\arcmin01.71\arcsec) with a diameter of $\rm\sim6.5\,kpc$\footnote{For comparison, the Small Magellanic Cloud has a diameter of 5.8 kpc and is it located 63 kpc from the Milky Way.}. H$\alpha$ emission is detected in the core of the galaxy, as well as in a small extended region to the East (see Figure~\ref{fig:MysteriousGalaxy}). The SN1, SN2, and SN3 spectra obtained after sky subtraction for this emission region were extracted with \texttt{ORCS} from an elliptical region with a semi-major axis of $\sim 875$ pc and a semi-minor axis of $\sim 375$ pc. The ellipse was selected to maximize the measured flux. The average heliocentric velocity of this region is $2635\,\kms$, well within the range displayed by NGC~2207 (2475 - $2900\,\kms$ -- see Figure~\ref{fig:Maps_sig-vel}). Furthermore, the absence of the [\ion{N}{ii}]$\lambda$6583 line, indicates the low metallicity of the region.
    
        This object resembles an edge-on dwarf spiral galaxy with a hint of a warped structure. With no obvious bridge seen in our images connecting this object to the interacting system, we infer that it is more likely distorted by the interaction, rather than a fragment of the larger galaxies ejected during the interaction. 
        
    \subsection{A dwarf spheroidal}
\begin{figure}
                \centering
                \includegraphics[width=\columnwidth]{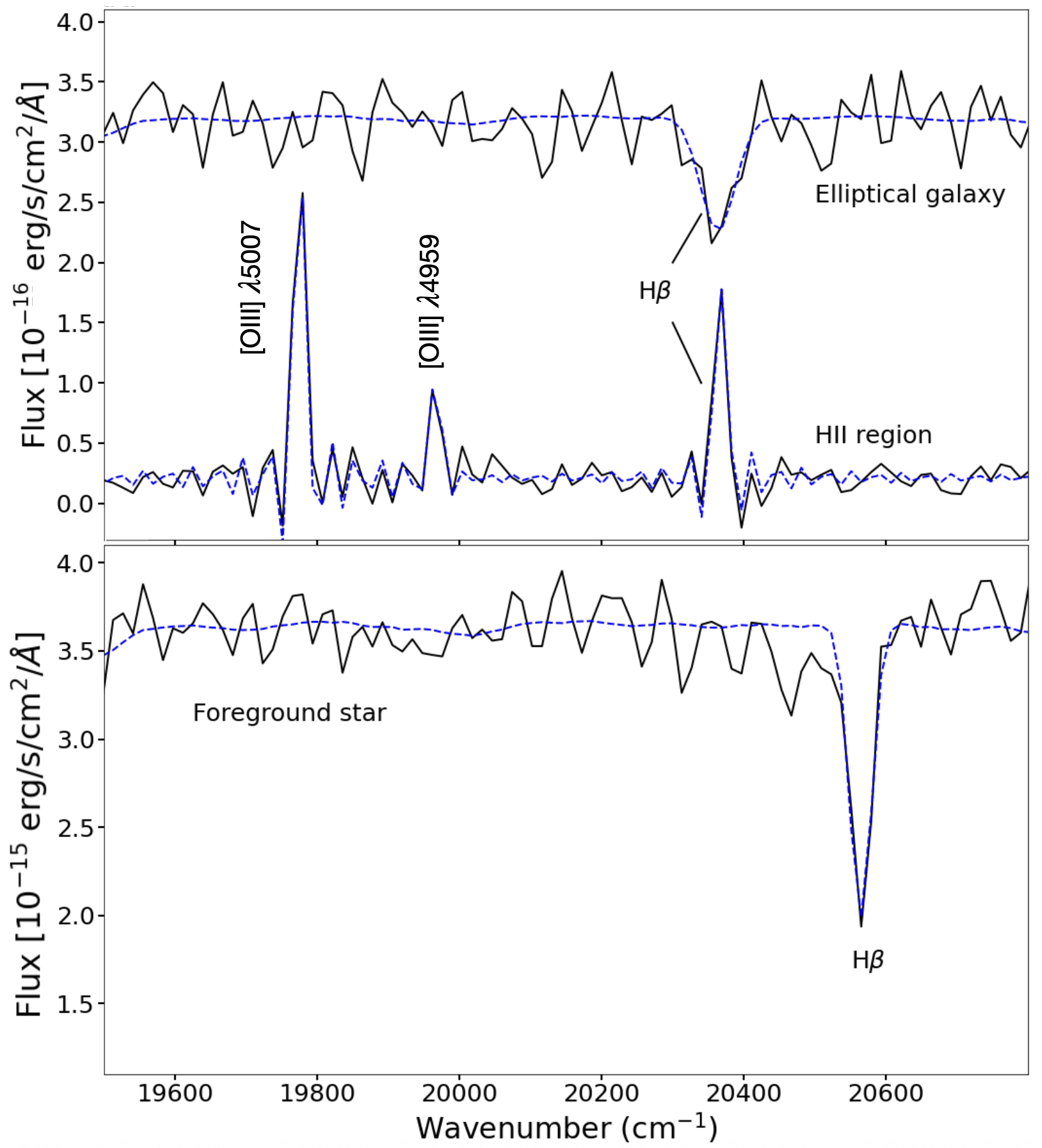}
                \caption{ (Top panel) Spectrum in the SN2 band of the dwarf elliptical galaxy at the tip of IC~2163's tidal tail and of a nearby \ion{H}{ii} region; (Lower panel) Spectrum of a field star in the Milky Way.  In all cases, the blue dashed line is a fit by ORCS.}
                \label{fig:DwarfElliptical}
            \end{figure}

        At the eastern tip of IC~2163's extended tidal tail lies a small round galaxy devoid of ionised gas (R.A. 06h16m35.80s, Dec. -21°22\arcmin02.7\arcsec) . Although it has been suggested to also be part of the interacting system \citep{elmegreen1995interactionI}, there have been to our knowledge no measurement of its systemic velocity. iFTS are not optimized for absorption spectroscopy of faint sources because of the distributed noise intrinsic to the technique \citep{araafts}, but we have demonstrated that SITELLE is nevertheless capable of detecting absorption features in stars and galaxies \citep{drissen2019sitelle,ruest6888}. The strong stellar lines characteristic of elliptical galaxies, such as Ca H and K, the G band, NaD or Mgb, are outside the spectral domain of the filters we used for this project. But we have attempted to extract the integrated spectrum of this galaxy using an aperture of radius 3$''$: the SN1 and SN3 data cubes do not show unambiguous stellar features, but the H$\beta$ absorption line is clearly detected in the SN2 cube (Figure~\ref{fig:DwarfElliptical}). We measure a systemic velocity of $(3005\pm65)\,\kms$, which is about $70\,\kms$ higher that of the nearby \ion{H}{ii} regions in IC~2163. Given its privileged location in the direct prolongation of IC~2163's tidal tail and its very similar systemic velocity, it is very tempting to suggest that this dwarf elliptical galaxy is not only physically related to this group, but also that it could fully participate in the interaction.
    
\section{Summary and Conclusions}
\label{section:Conclusions}

    Integral field spectroscopy of the NGC~2207/IC~2163 interacting pair obtained in the visible band with SITELLE has been presented. Our observations were combined with detailed numerical simulations of the interaction with GCD+. We highlight in the following the main results of this work.
    
        \begin{enumerate}
            \item More than 1100 star-formation complexes and \ion{H}{ii} regions are identified in the H$\alpha$ map and are used to determine the nebular abundances across both galaxies. Complementary methods were used to determine the host galaxy of 240 emission-line regions in the zone where both galaxies overlap; we found an excellent agreement between them, and the membership of only 22 regions remain ambiguous.
        
            \item The oxygen abundance gradients of \ion{H}{ii} region complexes were derived using four different strong-line indicators, calibrated from six different methods found in the literature. Despite differences in absolute values, all show a clear radial decline in both galaxies. With slope $\sim$$-$0.015 dex/kpc, the global O/H gradients are shallower than in isolated spirals in both galaxies, consistent with previous results for interacting systems. Some indicators suggest subtle discontinuities in the gradients but these are not universally seen. No significant azimuthal variations are detected within the general uncertainties of the strong-line indicators, suggesting that both systems are well mixed. Simulations reproduce these trends, with interaction-driven gas flows and localized SF accounting for the observed metallicity patterns.

            \item The H$\alpha$ luminosity functions of the \ion{H}{ii} regions are similar between both galaxies, while the bulk of regions in the arms of both galaxies has L(H$\alpha$) higher than $\simeq$ 0.4 dex compared to the inter-arm regions. In IC~2163, the slope of the LF seen for the regions located in the arms of the galaxy, found mostly in the eyelids, appears shallower than for the inter-arm regions, although the uncertainties are large due to the smaller sample of regions. This difference suggests that the molecular cloud mass spectrum in the eyelids differs due to the strong "piling up" of gas found in that region caused by the strong streaming motions generated during the collision. 
        
            \item Regions with large velocity dispersion are identified across the system, revealing diverse physical processes. In IC~2163, large $\sigma$ values (up to 65~km~s$^{-1}$) are observed near the northern eyelid, likely tracing turbulent gas stirred by streaming motions from the collision. In NGC~2207, the nucleus shows AGN-like emission with broad lines ($\sigma~\simeq~120$~km~s$^{-1}$), strong [\ion{N}{ii}]/H$\alpha$ and evidence of outflows. Outside some \ion{H}{ii} region complexes, diffuse ionised gas exhibits elevated dispersion (50 - 80~km~s$^{-1}$), likely tied to stellar feedback. Finally, \textit{Feature~i}, the brightest star-forming complex, displays enhanced $\sigma$ offset from its core, hinting at possible champagne flows or wind-driven turbulence. 
            
            \item Numerical simulations of the NGC~2207/IC~2163 interaction provide a powerful framework for exploring how galaxy encounters shape morphology, star formation and chemical enrichment. By modeling a grazing interaction between two spirals, the simulations reproduce the main observed features -- including tidal tails, ocular structures, and kinematic and metallicity patterns -- and reveal how close passages trigger starbursts and drive chemical evolution. In particular, they indicate a first pericentre passage about 160~Myr ago, when tidal forces funneled gas toward the galaxy centers, producing a starburst with a peak SFR of 10~M$_\odot$~yr$^{-1}$ and raising 12 + log(O/H) and, more gradually, [Fe/H] by 0.09 and 0.08~dex, respectively, between the first passage and the ``present''. These processes also explain the flattened oxygen abundance gradient in NGC~2207, the central metallicity enhancement in IC~2163, and the observed kinematic disturbances. Overall, the simulation underscore their value as diagnostic tools for disentangling the dynamical and physical processes shaping interacting galaxies.
        
            \item Two dwarf galaxies in the vicinity of the NGC~2207/IC~2163 pair are found to have very similar systemic velocities as their larger companions. Circumstantial evidence (the warped morphology of the northern star-forming galaxy, and the location of the dwarf elliptical at the very tip of IC~2163's tidal tail) suggest that they may participate in the interaction. They were not included in the numerical simulations.
        \end{enumerate}

\section*{Acknowledgements}
    Based on observations obtained with SITELLE, a joint project of Universit\'e Laval, ABB, Universit\'e de Montr\'eal, and the Canada-France-Hawaii Telescope (CFHT) which is operated by the National  Research Council of Canada, the Institut National des Sciences de  l'Univers of the Centre National de la Recherche Scientifique of France, and the University of Hawaii. The authors wish to recognize and acknowledge the very significant cultural role that the summit of Mauna Kea has always had within the indigenous Hawaiian community. We are most grateful to have the opportunity to conduct observations from this mountain. We are very grateful to David Rupke for providing unpublished spectrophotometric measurements of star-forming regions in NGC~2207/IC~2163. C.P. acknowledges support from the FRQNT master's training scholarship (https://doi.org/10.69777/346450). L.D., H.M., and C.R. are grateful to the Natural Sciences and Engineering Research Council of Canada, the Fonds de Recherche du Qu\'ebec (https://doi.org/10.69777/283645), and the Canada Foundation for Innovation for funding. R.P.M. is grateful to the Univ. of Hawaii at Hilo for the observing time allocation and support during the initial phase of this project. We thank the referee for valuable comments.

\section*{Data Availability Statement}
    Part of the data underlying this article are available in its online supplementary material; other data can be shared upon reasonable request to the authors.

\bibliographystyle{mnras}
\bibliography{references}

\appendix

    \section{Deep starless image}
    \label{appendix:Deepstarless}
    
        \begin{figure*}
            \centering
            \includegraphics[width=0.9\textwidth]{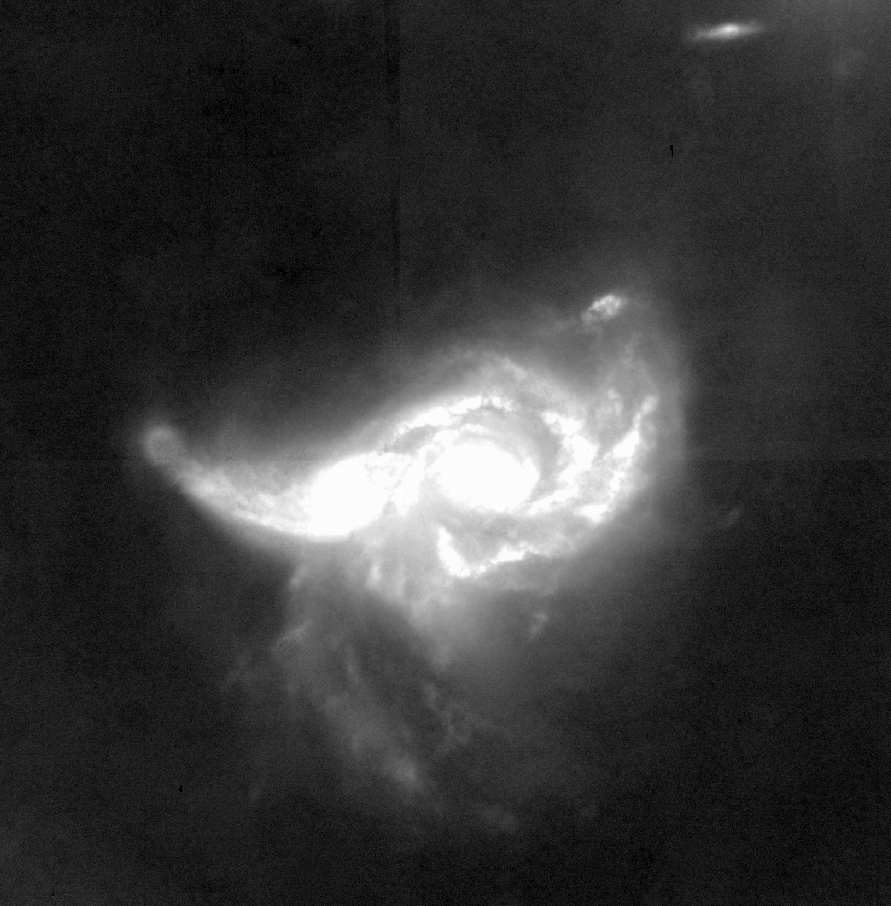}
            \caption{Deep {\it starless} image of the NGC~2207/IC~2163 pair from SITELLE, to illustrate the low surface brightness features in the outskirts of the galaxies. This image was obtained by first combining the deep frames from the SN1, SN2, and SN3 data cubes, then removing the point-like sources using the StarNet software (https://www.starnetastro.com/). Note that the core of the dwarf elliptical galaxy at the tip of IC~2163's tidal tail was unduly supressed in this process. Very low amplitude artefacts are also observed at the junction of the CCD quadrants.}
            \label{fig:DeepStarless}
        \end{figure*}

    \section{Comparison of emission line ratios}
    \label{appendix:ComparisonRupke}
    
        \begin{figure*}
            \centering
            \includegraphics[width=0.9\textwidth]{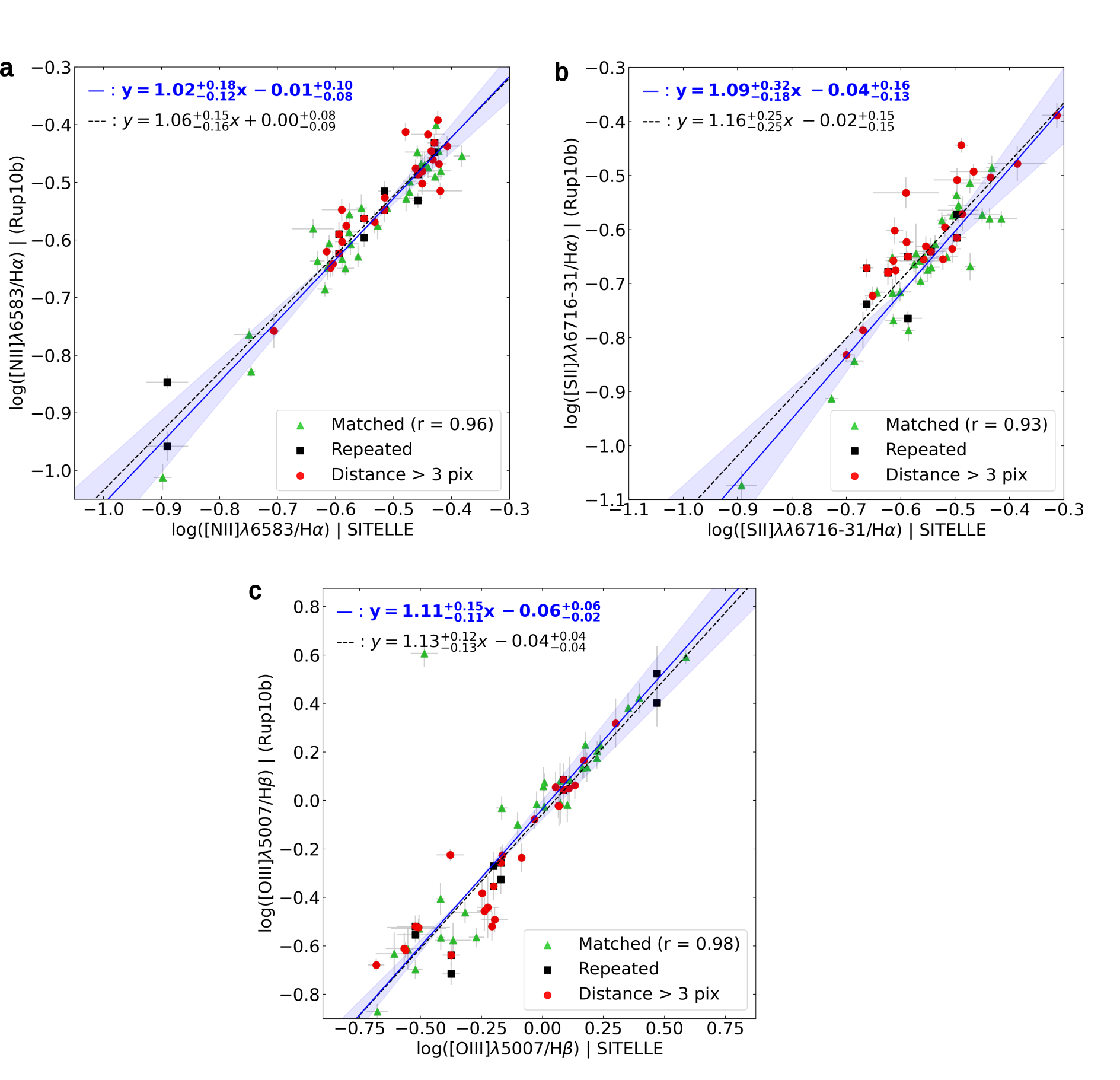}
            \caption{Comparison between the emission line flux ratios from \protect\citetalias{rupke2010gas} and SITELLE: (a)~[\ion{N}{ii}]$\lambda$6583/H$\alpha$, (b)~([\ion{S}{ii}]$\lambda$6716+[\ion{S}{ii}]$\lambda$6731)/H$\alpha$ and (c)~[\ion{O}{iii}]$\lambda$5007/H$\beta$. For all ratios, regions matched within a Euclidean distance of less than 3 spaxels are represented by green triangles, while those with a distance greater than 3 spaxels are shown as red circles. Regions identified by black squares correspond to SITELLE regions that were assigned to multiple \protect\citetalias{rupke2010gas} regions. The black dashed line represents the linear regression using all data points, whereas the solid blue line shows the linear regression using only the green triangles (matched regions). The Pearson correlation coefficient $r$ for these matched regions is indicated in the legend.}
            \label{fig:ComparisonRupke}
        \end{figure*}

    \section{BPT diagrams by H$\alpha$ and location}
    \label{appendix:BPTDiagrams-Threshold}
        Figure~\ref{appendix:BPTDiagrams-Threshold} presents the BPT diagrams split by an H$\alpha$ flux threshold to investigate how lower fluxes and higher uncertainties affect the dispersion, and by location to assess possible differences between the two galaxies and between the arm and inter-arm regions (see Section~\ref{section:Separation} and Section~\ref{section:LuminosityFunctions} for the respective definitions).
    
            \begin{figure*}
                \centering
                \includegraphics[width=\textwidth]{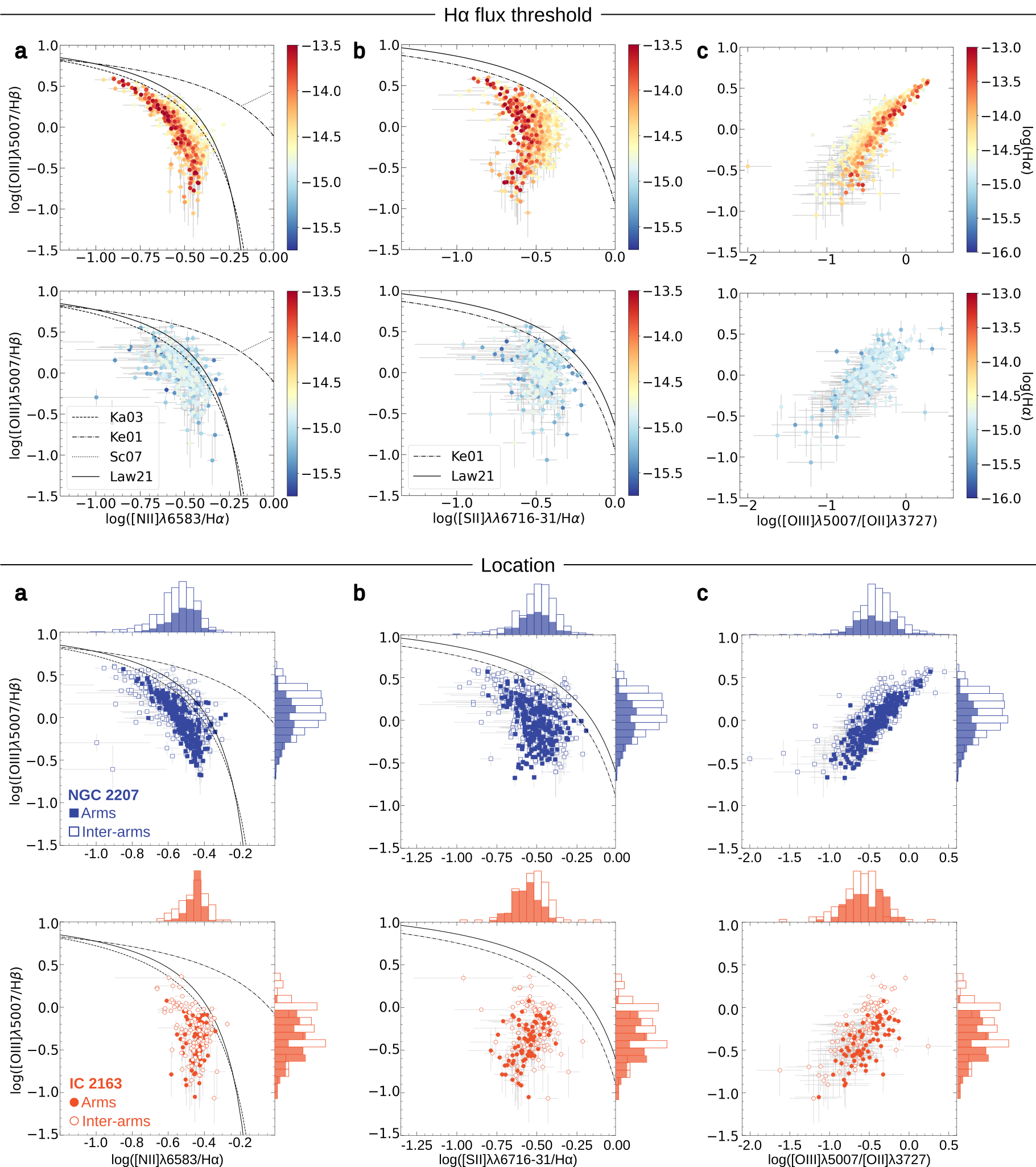}
                \caption{BPT diagrams for \ion{H}{ii} region complexes displaying the [\ion{O}{iii}]/H$\beta$ ratios as a function of (a)~[\ion{N}{ii}]/H$\alpha$, (b)~[\ion{S}{ii}]/H$\alpha$, and (c)~[\ion{O}{iii}]/[\ion{O}{ii}]. In the top panel group, complexes are separated by H$\alpha$ flux: the upper panels correspond to $\log{(\text{H}\alpha)} \geq -14.75$, while the lower panels correspond to values below this threshold. The dispersion is more pronounced for lower H$\alpha$ fluxes due to higher uncertainties. In the bottom panel group, complexes are separated by location: panels show regions belonging to NGC~2207 (blue) and IC~2163 (orange), with arms and inter-arms shown as filled and unfilled symbols, respectively. For each diagram, histograms in the margins (top and right of each panel) show the distributions of the corresponding line ratios along the $x$ and $y$ axes. Each pair of histograms (arms vs inter-arms) is scaled relative to the maximum count within the pair. In all diagrams, classification boundaries are indicated following \protect\citet[dashed line]{kauffmann2003host}, \protect\citet[dash-dotted line]{kewley2001optical}, \protect\citet[dotted line]{schawinski2007observational}, and \protect\citet[solid line]{law2021sdss}.}
                \label{fig:BPTDiagrams-Threshold}
            \end{figure*}

    \section{Velocity map of the simulated gas component}
        Although the simulated gas and stellar component exhibit globally similar LOS velocity fields, the gas velocity map contains larger regions with sparse sampling and mixed velocities, particularly in the low-density outskirts and tidal features. This effect arises from the lower number of gas particles compared to stellar component.
        
            \begin{figure*}
                \centering
                \includegraphics[width=\textwidth]{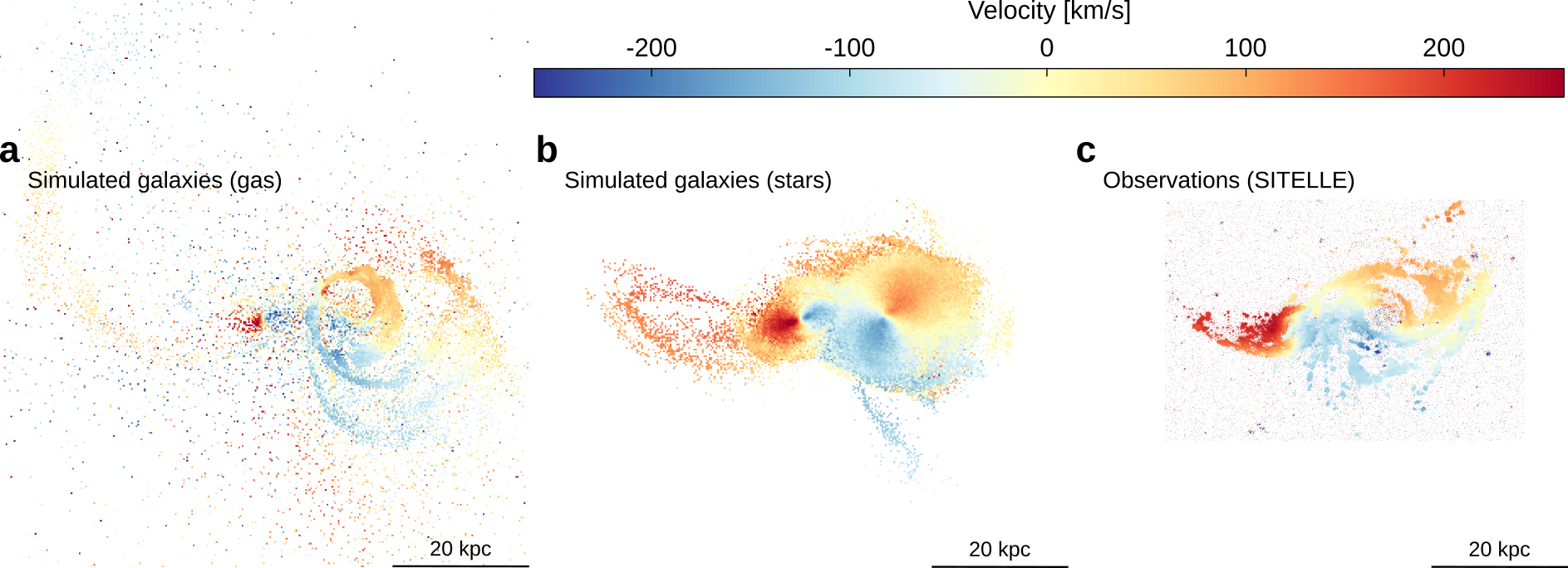}
                \caption{LOS velocity maps of the interacting system at the “present” time snapshot ($t = 400$~Myrs). Panels show the simulated (a)~gas component and (b)~stellar component. (c)~Observed LOS velocity map obtained with SITELLE, shown relative to a systemic velocity of $2750$~km~s$^{-1}$. The same color scale is applied to all panels. Spatial scales are indicated in each panel.}
                \label{fig:KinematicsSimulation-gas}
            \end{figure*}

    \section{Properties of the HII region complexes}

        \begin{table*}
            \centering
            \footnotesize
            \caption{Excerpt from the catalog of \ion{H}{ii} region complexes. The full table is available in the online version.}
            \label{tab:properties_complexes}
        
            \begin{tabular}{rcccccccc}
                \hline
                ID & R.A. (J2000) & DEC (J2000) & Galaxy & $F(\text{H}\alpha)$ & $F($[\ion{S}{II}]$\lambda6731)$ & $F($[\ion{S}{II}]$\lambda6716)$ & $F($[\ion{N}{II}]$\lambda6583)$ & $F($[\ion{N}{II}]$\lambda6548)$ \\
                \hline
                1 & 06:16:22.90 & -21:24:54.5 & N. & 3.92(0.46)  & 0.31(0.33) & 0.46(0.33) & 0.50(0.32) & -0.03(0.31) \\
                2 & 06:16:23.79 & -21:24:09.5 & N. & 25.09(1.04) & 3.35(0.75) & 5.45(0.76) & 6.16(0.74) & 1.37(0.72) \\
                3 & 06:16:25.27 & -21:23:46.3 & N. & 12.39(0.69) & 1.29(0.50) & 2.13(0.51) & 3.90(0.51) & 0.18(0.48) \\
                4 & 06:16:20.00 & -21:23:46.9 & N. & 5.51(0.50)  & 1.14(0.37) & 0.57(0.36) & 1.12(0.35) & 1.32(0.35) \\
                5 & 06:16:19.42 & -21:23:46.4 & N. & 22.84(0.63) & 1.77(0.45) & 2.35(0.45) & 3.57(0.44) & 1.94(0.43) \\
                \multicolumn{9}{c}{\dots} \\
                \hline
            \end{tabular}
        
            \vspace{8pt}
        
            \begin{tabular}{rcccccc}
                \hline
                ID & $F($[\ion{O}{III}]$\lambda5007)$ & $F($[\ion{O}{III}]$\lambda4959)$ & $F(\text{H}\beta)$ & $F($[\ion{O}{II}]$\lambda3727)$ & $v$ & $\sigma$ \\
                \hline
                1 & 1.96(0.57)  & 0.53(0.43)  & 1.37(0.49) & 4.41(3.35) & 2668.5(4.1) & 23.3(9.1) \\
                2 & 17.72(0.99) & 7.01(0.76) & 8.74(0.78) & 26.54(5.69) & 2711.9(1.5) & 27.3(2.8) \\
                3 & 7.59(0.99)  & 3.90(0.80) & 4.32(0.81) & 14.51(3.97) & 2695.2(1.8) & 19.7(4.9) \\
                4 & 1.87(0.38)  & -0.15(0.37) & 1.92(0.36) & 7.86(3.82) & 2655.4(3.0) & 20.0(7.9) \\
                5 & 23.61(0.72) & 7.90(0.53) & 7.96(0.52) & 22.28(4.38) & 2651.9(1.0) & 23.0(2.2) \\
                \multicolumn{7}{c}{\dots} \\
                \hline
            \end{tabular}
            \par\vspace{1ex}
            \parbox{0.9\linewidth}{\footnotesize Note : The ID corresponds to the index of the complex in the catalog, along with its coordinates. The column 'Galaxy' indicates the host galaxy assignment: N. for NGC~2207, I. for IC~2163 and Amb.\ for ambiguous cases. Fluxes $F$ for each specified line are in units of $10^{-16}$ erg s$^{-1}$ cm$^{-2}$, corrected for extinction; the radial velocity $v$ and velocity dispersion $\sigma$ in km\,s$^{-1}$.}
        \end{table*}

        \begin{table*}
            \centering
            \footnotesize
            \caption{Excerpt from the catalog of the marginal detections. The full table is available in the online version.}
            \label{tab:properties_marginal}
        
            \begin{tabular}{rccccccc}
                \hline
                ID & R.A. (J2000) & DEC (J2000) & $F(\text{H}\alpha)$ & $F($[\ion{S}{II}]$\lambda6731)$ & $F($[\ion{S}{II}]$\lambda6716)$ & $F($[\ion{N}{II}]$\lambda6583)$ & $F($[\ion{N}{II}]$\lambda6548)$ \\
                \hline
                1-marg & 06:16:25.30 & -21:25:00.3 & 2.66(0.43) & 1.24(0.36) & 1.13(0.36) & 1.14(0.35) & 0.69(0.33) \\
                2-marg & 06:16:19.53 & -21:24:54.9 & - & - & - & - & - \\
                3-marg & 06:16:19.55 & -21:24:53.9 & - & - & - & - & - \\
                4-marg & 06:16:23.71 & -21:24:52.5 & - & - & - & - & - \\
                5-marg & 06:16:21.96 & -21:24:49.2 & 2.49(0.24) & -0.02(0.25) & -0.33(0.25) & 0.61(0.24) & 0.14(0.24) \\
                \multicolumn{8}{c}{\dots} \\
                \hline
            \end{tabular}
        
            \vspace{8pt}
        
            \begin{tabular}{rcccccc}
                \hline
                ID & $F([\text{OIII}]\lambda5007)$ & $F([\text{OIII}]\lambda4959)$ & $F(\text{H}\beta)$ & $F([\text{OII}]\lambda3727)$ & $v$ & $\sigma$ \\
                \hline
                1-marg & 1.21(0.27) & 1.03(0.27) & 0.53(0.26) & 4.72(4.51) & 2683.3(5.1) & 25.3(10.8) \\
                2-marg & - & - & - & - & - & - \\
                3-marg & - & - & - & - & - & - \\
                4-marg & - & - & - & - & - & - \\
                5-marg & 2.08(0.43) & 1.11(0.35) & 1.36(0.36) & 5.76(3.56) & 2654.0(4.4) & - \\
                \multicolumn{7}{c}{\dots} \\
                \hline
            \end{tabular}
             \par\vspace{1ex}
            \parbox{0.9\linewidth}{\footnotesize Note :  The ID corresponds to the index of the complex in the catalog, along with its coordinates. Fluxes $F$ for each specified line are in units of $10^{-16}$ erg s$^{-1}$ cm$^{-2}$; the radial velocity $v$ and velocity dispersion $\sigma$ in km\,s$^{-1}$.}
        \end{table*}

\bsp	
\label{lastpage}
\end{document}